\documentclass[11pt]{article}
\usepackage{amssymb}

\usepackage{curves}
\usepackage{amsmath}

\newcounter{resultnum}[section]

\newcounter{conclusionnum}[section]

\newcounter{conditionnum}[section]

\newcounter{conjecturenum}[section]

\newcounter{examplenum}[section]

\newcounter{exercisenum}[section]

\newcounter{lemmanum}[section]

\newcounter{notationnum}[section]

\newcounter{theoremnum}[section]

\newcounter{definitionnum}[section]

\newcounter{corollarynum}[section]

\newcounter{remarknum}[section]

\newcounter{propositionnum}[section]

\newcounter{acknowledgementnum}[section]

\newcounter{algorithmnum}[section]

\newcounter{axiomnum}[section]

\newcounter{casenum}[section]

\newcounter{claimnum}[section]

\newcounter{summarynum}[section]

\newcounter{problemnum}[section]

\begin{document}

\title{Loop Quantum Gravity in Ashtekar and Lagrange--Finsler Variables and
Fedosov Quantization of General Relativity }
\date{February 9, 2008}
\author{ Sergiu I. Vacaru\thanks{%
sergiu$_{-}$vacaru@yahoo.com, svacaru@fields.utoronto.ca } \\
\textsl{\small The Fields Institute for Research in Mathematical Science} \\
\textsl{\small 222 College Street, 2d Floor, Toronto, Ontario  M5T 3J1, Canada}
\\ {\small and}\\
\textsl{\small Perimeter Institute for Theoretical Physics}\footnote{short term visit}\\
\textsl{\small Waterloo, Ontario N2J 2W9, Canada}
}
\maketitle

\begin{abstract}
We propose an unified approach to loop quantum gravity and Fedosov
quantization of gravity following the geometry of double spacetime
fibrations and their quantum deformations. There are considered
pseudo--Riemannian manifolds enabled with 1) a nonholonomic 2+2 distribution
defining a nonlinear connection (N--connection) structure and 2) an
Arnowitt--Deser--Misner 3+1 decomposition. The Ashtekar--Barbero variables
are generalized and adapted to the N--connection structure which allows us
to write the general relativity theory equivalently in terms of
Lagrange--Finsler variables and related canonical almost symplectic forms
and connections. The Fedosov results are re--defined for gravitational gauge
like connections and there are analyzed the conditions when the star product
for deformation quantization is computed in terms of geometric objects in
loop quantum gravity. We speculate on equivalence of quantum gravity
theories with 3+1 and 2+2 splitting and quantum analogs of the Einstein
equations.

\vskip5pt

\textbf{Keywords:}\ Loop quantum gravity, nonlinear connections, general
relativity, nonholonomic manifolds, nonholonomic frames, Ashtekar--Barbero
variables, Fedosov quantization, deformation quantization, Finsler geometry
methods, (pseudo) Riemannian and Lagrange spaces.

\vskip10pt

MSC:\ 83C45, 83C99, 53D55, 53B40, 53B35

PACS:\ 04.60.Pp, 02.40.-k, 02.90.+g, 03.65.Sq 
\end{abstract}

\tableofcontents


\section{ Introduction}

In recent thirty years, there were developed three most popular approaches
to quantization of gravity and unified and nonlinear theories in physics:
the string/M--theory (SMT), geometric quantization and deformation
quantization (DQ) and loop quantum gravity (LQG).

The SMT aims to a theory unifying all (in general, supersymmetric)
interactions on higher dimension spacetimes and proposes a general physical
paradigm containing as particular cases the well known four dimensional
classical and quantum models of gravity and matter field interactions, see
reviews of results and methods in Refs. \cite{str1,str2,str3}.\footnote{%
in this article, we do not aim to outline and discuss the achievements of
string theory or to present comprehensive bibliographies on this and
below--mentioned directions; for more details, see cited works and
references therein}

The second approach (we cite here some important works on DQ \cite%
{defq1,defq2,fed1,fed2,konts1,karabeg1,sternh1,ali}) concerns various
scenarios of geometric quantization of linear and nonlinear field theories
when there are synthesized and developed the methods of modern symplectic/
Poisson geometry, algebroid mathematics, quantum group theory and
noncommutative geometry \cite{vais,acs,maj,connes,madore,gbvf}.

In contrast to very general purposes to unification of physical interactions
and elaborating general geometric principles of quantization (respectively,
in SMT and DQ), the third approach, i.e. LQG, was performed just as a theory
of quantum gravity combining the general relativity (GR) and quantum
mechanics, see reviews of results in Refs. \cite{rov,thiem1,asht,smol}. This
implies a non--perturbative formulation when the background independence
(the key feature of Einstein's theory) is preserved. At present time, LQG is
supposed to have a clear conceptual setup following from physical
considerations and supported by a rigorous mathematical formulation. The
existing criticism against LQG (see, for instance, \cite{nicolai}) is
considered to be motivated in the bulk by arguments that the mathematical
formalism is not that which is familiar for the particle physicists working
with perturbation theory, Fock spaces, background fields etc, see details
and discussion \cite{thiem}.

In a series of our works \cite{esv,vqg1,vqg2,vqg3,vqg4,avqg5}, we developed
the idea that it is possible to re--formulate equivalently the GR in a
formal language of Lagrange--Finsler variables, with a nonholonomic
spacetime 2+2 splitting defining canonically an almost K\"{a}hler geometry.%
\footnote{%
We note that we shall work with (pseudo) Riemannian spaces enabled with
nonholonomic distributions modelling effective Lagrange--Finsler geometries
on nonholonomic manifolds, i.e. with certain classes of Lagrange--Finsler
variables in GR but not on tangent bundles; see details on physical
applications of Finsler geometry methods in standard models of physics in
Refs. \cite{vrfg,vsgg,vncg,vijgmmp}.} That allowed us to develop some models
of Fedosov quantization for the Einstein gravity (with violation of local
Lorentz symmetry \cite{vqg2,avqg5}, or preserving this symmetry \cite%
{vqg3,vqg4}, certain generalizations to Lagrange--Finsler \cite%
{esv,vqg1,vqg2} and Hamilton--Cartan \cite{avqg5} spaces and analogous
modelling by Lagrange and/or Finsler systems of interactions in classical
and quantum gravity).\footnote{%
Former works on deformation quantization did not concern an explicit
quantization of GR because in the past there were not constructed natural
almost symplectic variables (2-forms and connections) which would be
equivalent to the (pseudo) Riemannian metric and related Levi Civita
connection. Nevertheless, a number of interesting approaches to deformation
quantizations and gauge gravity models, generalized gravity and linearized
versions were considered, see discussions and original results in Refs. \cite%
{gcpp,ant,quev,castro}.} Using nonholonomc and quantum deformations on
(pseudo) Riemannian, or effective (induced by deformations of certain
geometric structures) Riemann--Cartan manifolds, the procedure of
quantization can be performed in certain forms preserving the Lorentz local
invariance. If the quantization formalism is developed on (co) tangent
bundles, one gets quantum corrections violating this local symmetry. The
Lagrange--Finsler variables associated to 2+2 distributions are similar to
the Ashtekar--Barbero \cite{asht1,asht2,barb1,barb2} variables and related
spacetime\ 3+1 splitting (details on the so--called ADM, i.e.
Arnowitt--Deser--Misner, formalism in GR can be found, for instance, in \cite%
{mtw,rov,thiem1,asht}). Both types of such variables (and other various
ones, like spinor variables and spin connections, tetradic variables etc)
can be equivalently introduced on (pseudo) Riemannian spacetimes and
generalized, for instance, on Riemann--Cartan spaces \cite%
{mercuri1,vrfg,vsgg}. The main difference is that the 2+2 splitting is more
convenient for certain methods and purposes of deformation quantization but
the 3+1 splitting, with shift and lapse functions related to a corresponding
Hamilton formalism, was chosen as the starting point for LQG. It should be
noted here that both approaches can be adapted to preserve the
diffeomorphism invariance and encode classical and quantum gravitational
interactions, non--perturbatively, into certain nonlinear geometric
structures.

The goal of this article is to analyze the conditions when two types of
quantization of gravity, i.e. the LQG and DQ, admit mutual transforms of
geometric and physical objects and fundamental equations. The task to
establish certain equivalence conditions for two quantizations of generic
nonlinear theories is not trivial: The 'deformation philosophy' \cite%
{sternh1} is very different from the paradigm of LQG and the mathematical
methods and elaborated schemes of quantization belong to quite different
branches of nonlinear functional analysis and almost Hermitian geometry. Our
intention is to compare the mentioned formalisms of quantum gravity and give
self--consistent and complimentary points of view together with a synthesis
of ideas outlined in separate forms in \cite{rov,thiem1,asht,smol} and \cite%
{defq1,defq2,fed1,fed2,konts1,karabeg1,sternh1,ali}. This is the first paper
in a series of works (the second partner is \cite{vlqgdq2}) oriented to a
comparative study of the methods and results of LQG and DQ. We shall also
consider possible applications of quantum gravity techniques in geometric
mechanics, string theory and noncommutative geometry and analogous gravity
for modelling respective quantum and classical interactions.

The paper is organized as follows:

In section 2, we formulate the general relativity theory in terms of
Lagrange-Finsler variables which allows us to define certain canonical
nonlinear connection and almost complex/ symplectic structures induced by
(pseudo) Riemannian metrics. The Einstein gravity is formulated equivalently
on almost K\"{a}hler spaces.

Section 3 is devoted to a (3+1)--splitting formalism in general relativity
which is adapted to a fixed nonlinear connection structure (induced
effectively by certain off--diagonal metric terms). The corresponding
Hamiltonian formulations and generalized phase constructions are provided.

In section 4, we generalize the Ashtekar connections in a form allowing
straightforward applications of deformation quantization methods formally
elaborated for almost symplectic spaces enabled with a nontrivial Neijenhuis
structure and compatible affine connection. We compute the star product in
Lagrange--Finsler variables in general relativity and define the
Fedosov--Ashtekar distinguished operators.

We provide the main results on deformation quantization of gravity in terms
of nonholonomically generalized Ashtekar connections in section 5. There are
also computed the Chern--Weyl form defining the zero--degree cohomology for
Einstein spaces and discussed the problem of quantum deformation imbedding
of vacuum Einstein equations.

We conclude and discuss the results in section 6. In Appendix, there are
considered some necessary formulas on nonholonomic deformation of
connections and geometric objects.

\section{Lagrange--Finsler and Almost K\"{a}hler Formulations of Einstein
Gravity}

In this section we introduce two different classes of variables that are
used for the Lagrange--Finsler modelling of general relativity, GR, and
nonholonomic deformation quantization of GR and, respectively, in the
definition of loop quantum gravity, LQG. We emphasize that in both cases the
covariant character of four dimensional spacetime is preserved.

Before starting with the more technical material, let us establish some
important coordinate and index conventions:

We consider a real four dimensional (pseudo) Riemanian spacetime manifold $V$
of signature $(-,+,+,+)$ and necessary smooth class. For a conventional $2+2$
splitting, the local coordinates $u=(x,y)$ on a open region $U\subset V$ are
labelled in the form $u^{\alpha }=(x^{i},y^{a}),$ where indices of type $%
i,j,k,...=1,2$ and $a,b,c...=3,4,$ for tensor like objects, will be
considered with respect to a general (non--coordinate) local basis $%
e_{\alpha }=(e_{i},e_{a}).$ One says that $x^{i}$ and $y^{a}$ are
respectively the conventional horizontal/ holonomic (h) and vertical /
nonholonomic (v) coordinates (both types of such coordinates can be time--
or space--like ones). In the case of $3+1$ splitting, we label the
coordinates as $u^{\alpha }=(u^{0},u^{I}),$ where the space like indices run
values as $I,J,K...=1,2,3$ and $\mathring{0}$ is the time like index. We
shall use both types of indices as abstract or coordinate ones but underline
them, $\underline{i},\underline{a},\underline{I},\underline{J},...,$ in
order to emphasize that such indices are just coordinate ones.

Primed indices of type $i^{\prime },a^{\prime },I^{\prime },J^{\prime },...$
will be used for labelling coordinates with respect to a different local
basis $e_{\alpha ^{\prime }}=(e_{i^{\prime }},e_{a^{\prime }})$ or $%
e_{\alpha ^{\prime }}=(e_{0}^{\prime },e_{I^{\prime }}),$ for instance, for
an orthonormalized basis. For the local tangent Minkowski space, we chose $%
e_{0^{\prime }}=i\partial /\partial u^{0^{\prime }},$ where $i$ is the
imaginary unity, $i^{2}=-1,$ and write $e_{\alpha ^{\prime }}=(i\partial
/\partial u^{0^{\prime }},\partial /\partial u^{1^{\prime }},\partial
/\partial u^{2^{\prime }},\partial /\partial u^{3^{\prime }}).$ To consider
such formal Euclidean coordinates is useful for some purposes of analogous
modelling of gravity theories as effective Lagrange mechanics geometries,
but this does not mean that we introduce any complexification of classical
spacetimes.

We also note that Einstein's rule on summing up/low indices will be applied
in this work if the contrary will be not stated as a particular case. The
symbol ''$\doteqdot "$ means ''by definition". Boldface letters will be used
for spaces and geometric objects enabled/ adapted with/to certain
nonholonomic distributions with effective nonlinear connection structure on
a (pseudo) Riemannian spacetime $\mathbf{V},$ see our conventions from \cite%
{vsgg,vrfg,vqg1,vqg2,vqg3,vqg4,avqg5}.

\subsection{Lagrange--Finsler variables in GR}

Let us parametrize the coefficients of a general (pseudo) Riemannian metric
on a spacetime $V$ in the form: \
\begin{eqnarray}
\mathbf{g} &=&g_{i^{\prime }j^{\prime }}(u)e^{i^{\prime }}\otimes
e^{j^{\prime }}+h_{a^{\prime }b^{\prime }}(u)e^{a^{\prime }}\otimes
e^{b^{\prime }},  \label{gpsm} \\
e^{a^{\prime }} &=&\mathbf{e}^{a^{\prime }}-N_{i^{\prime }}^{a^{\prime
}}(u)e^{i^{\prime }},  \notag
\end{eqnarray}%
where the required form of vierbein coefficients $e_{\ \alpha }^{\alpha
^{\prime }}$ of the dual basis
\begin{equation}
e^{\alpha ^{\prime }}=(e^{i^{\prime }},e^{a^{\prime }})=e_{\ \alpha
}^{\alpha ^{\prime }}(u)du^{\alpha },  \label{dft}
\end{equation}%
defining a formal $2+2$ splitting, will be stated below.

\subsubsection{Effective Lagrangians for (pseudo) Riemannian metrics}

We consider an effective regular Lagrangian (fundamental, or generating,
function) $L(u)=L(x^{i},y^{a}),$ on a spacetime $V,$ defining the
nondegenerate Hessian\footnote{%
we get the so--called Lagrange metric if $y^{a}$ are associated to vertical
coordinates in a Lagrange geometry \cite{ma,avqg5} constructed on a tangent
bundle $TM$ with a base manifold $M$ labelled by coordinates $x^{i};$ we
emphasize here that regular Lagrange systems can be effectively modelled on
nonholonomic manifolds, i.e. manifolds enabled with nonholonomic
distributions defined canonically by a Lagrangian $L,$ see details in \cite%
{vrfg,vsgg}}
\begin{equation}
\ ^{L}h_{ab}=\frac{1}{2}\frac{\partial ^{2}L}{\partial y^{a}\partial y^{b}},
\label{elm}
\end{equation}%
when $\det |\ ^{L}h_{ab}|\neq 0,$ and
\begin{equation}
\ ^{L}N_{i}^{a}=\frac{\partial G^{a}}{\partial y^{2+i}},  \label{clnc}
\end{equation}%
for
\begin{equation}
G^{a}=\frac{1}{4}\ ^{L}h^{a\ 2+i}\left( \frac{\partial ^{2}L}{\partial
y^{2+i}\partial x^{k}}y^{2+k}-\frac{\partial L}{\partial x^{i}}\right) ,
\label{clssp}
\end{equation}%
where $\ ^{L}h^{ab}$ is inverse to $\ ^{L}h_{ab}$ and respective
contractions of $h$-- and $v$--indices, $\ i,j,...$ and $a,b...,$ are
performed following the rule: we can write, for instance, an up $v$--index $%
a $ as $a=2+i$ and contract it with a low index $i=1,2.$ Briefly, we shall
write $y^{i}$ instead of $y^{2+i},$ or $y^{a}.$ The values (\ref{elm}), (\ref%
{clnc}) and (\ref{clssp}) allows us to construct a Lagrange--Sasaky type
metric (we adapt for pseudo--Riemannian manifold of even dimension the
terminology used, for instance, in Refs. \cite{ma,vrfg})
\begin{eqnarray}
^{L}\mathbf{g} &=&\ ^{L}g_{ij}dx^{i}\otimes dx^{j}+\ ^{L}h_{ab}\ ^{L}\mathbf{%
e}^{a}\otimes \ ^{L}\mathbf{e}^{b},  \label{lfsm} \\
\ ^{L}\mathbf{e}^{a} &=&dy^{a}+\ ^{L}N_{i}^{a}dx^{i},\ ^{L}g_{ij}=\
^{L}h_{2+i\ 2+j}.  \notag
\end{eqnarray}

Let us consider a subclass of frame transforms $e^{\alpha ^{\prime }}=e_{\
\alpha }^{\alpha ^{\prime }}du^{\alpha },$ when the coefficients are
parametrized in the form
\begin{equation*}
e^{i^{\prime }}=e_{\ i}^{i^{\prime }}dx^{i},e^{a^{\prime }}=e_{\
i}^{a^{\prime }}dx^{i}+e_{\ a}^{a^{\prime }}dy^{a},
\end{equation*}%
with $e_{\ a}^{i^{\prime }}=0.$ A metric $\mathbf{g}$ (\ref{gpsm}) with
coefficients $g_{\alpha ^{\prime }\beta ^{\prime }}=[g_{i^{\prime }j^{\prime
}},h_{a^{\prime }b^{\prime }}]$ computed with respect to a dual basis $%
e^{\alpha ^{\prime }}=(e^{i^{\prime }},e^{a^{\prime }})$ is equivalent to a
metric $^{L}\mathbf{g}$ (\ref{lfsm}) with coefficients $\ ^{L}\mathbf{g}%
_{\alpha \beta }=\ [\ ^{L}g_{ij},\ ^{L}h_{ab}]$ defined with respect to a
N--adapted dual basis $\ ^{L}e^{\alpha }=(dx^{i},\ ^{L}\mathbf{e}^{a})$ if
there are satisfied the conditions%
\begin{equation*}
\mathbf{g}_{\alpha ^{\prime }\beta ^{\prime }}e_{\ \alpha }^{\alpha ^{\prime
}}e_{\ \beta }^{\beta ^{\prime }}=~^{L}\mathbf{g}_{\alpha \beta }.
\end{equation*}%
Considering any given values $\mathbf{g}_{\alpha ^{\prime }\beta ^{\prime }}$
and $~^{L}\mathbf{g}_{\alpha \beta }$ induced by a generating function $%
L(x,y),$ we have to solve a system of quadratic algebraic equations with
unknown variables $e_{\ \alpha }^{\alpha ^{\prime }}.$ For instance, in GR,
there are 6 independent values $\mathbf{g}_{\alpha ^{\prime }\beta ^{\prime
}}$ and up till ten coefficients $~^{L}\mathbf{g}_{\alpha \beta }$ which
allows us always to define a set of vierbein coefficients $e_{\ \alpha
}^{\alpha ^{\prime }};$ for many cases, a subset of such coefficients can be
taken be zero. Usually, we can use the N--adapted system of equations:
\begin{eqnarray}
g_{i^{\prime }j^{\prime }}e_{\ i}^{i^{\prime }}e_{\ j}^{j^{\prime
}}+h_{a^{\prime }b^{\prime }}e_{\ i}^{a^{\prime }}e_{\ j}^{b^{\prime }} &=&\
^{L}g_{ij}+\ ^{L}h_{ab}\ ^{L}N_{i}^{a}\ ^{L}N_{j}^{b},  \label{algeq1} \\
h_{a^{\prime }b^{\prime }}e_{\ i}^{a^{\prime }}e_{\ b}^{b^{\prime }} &=&\
^{L}h_{ab}\ ^{L}N_{i}^{a},\quad  \label{algeq2} \\
h_{a^{\prime }b^{\prime }}e_{\ a}^{a^{\prime }}e_{\ b}^{b^{\prime }} &=&\
^{L}h_{ab}.  \label{algeq3}
\end{eqnarray}%
We argue that a metric $\mathbf{g}$ is written in Lagrange variables on an
open region $U\subset \mathbf{V}^{2+2},$ induced by a generating function $%
L, $ if and only if there is a nontrivial real solution $e_{\ \alpha
}^{\alpha ^{\prime }}=[e_{\ i}^{i^{\prime }},e_{\ a}^{i^{\prime }}=0,e_{\
i}^{a^{\prime }},e_{\ b}^{b^{\prime }}]$ on $U$ of the system of quadratic
algebraic equations (\ref{algeq1})--(\ref{algeq3}), for given values $%
[g_{i^{\prime }j^{\prime }},h_{a^{\prime }b^{\prime }},N_{i^{\prime
}}^{a^{\prime }}]$ and $\ [\ ^{L}g_{ij},\ ^{L}h_{ab},\ \ ^{L}N_{i}^{a}],$
when
\begin{equation}
N_{i^{\prime }}^{a^{\prime }}=e_{i^{\prime }}^{\ i}e_{\ a}^{a^{\prime }}\ \
^{L}N_{i}^{a}  \label{defcncl}
\end{equation}%
for $e_{i^{\prime }}^{\ i}$ being inverse to $e_{\ i}^{i^{\prime }}.$

For simplicity, in this work, we suppose that there is always a finite
covering of $\mathbf{V}^{2+2}$ (in brief, denoted $\mathbf{V})$ by a family
of open regions $\ ^{I}U,$ labelled by an index $I,$ on which there are
considered certain nontrivial effective Lagrangians $\ ^{I}L$ with real
solutions $\ ^{I}e_{\ \alpha }^{\alpha ^{\prime }}$ defining vielbein
transforms to systems of Lagrange variables. It should be noted here that
the generating functions $\ ^{I}L$ are arbitrary ones on a $\ ^{I}U$ which
satisfy the conditions (\ref{elm}): we compute explicitly an effective$\
^{I}L$ by integrating two times a convenient for our purposes $\ ^{L}h_{ab}$
together with a prescribed $h_{a^{\prime }b^{\prime }}$ both resulting in a
real solution $\ ^{I}e_{\ a}^{a^{\prime }}$ of (\ref{algeq1}). Having any $\
^{I}L,$ $\ ^{L}h_{ab}$ and $\ ^{I}e_{\ a}^{a^{\prime }},$ we compute $%
^{L}N_{i}^{a}$ (\ref{clnc}) which allows us to define $\ ^{I}e_{\
i}^{a^{\prime }}$ from (\ref{algeq2}) (we have to change the partition $\
^{I}U$ and generating function $\ ^{I}L$ till we are able to construct real
solutions). Finally, we solve the algebraic equations (\ref{algeq1}) for any
prescribed values $g_{i^{\prime }j^{\prime }}$ (we also have to change the
partition $\ ^{I}U$ and generating function $\ ^{I}L$ till we are able to
construct real solutions) and find $\ ^{I}e_{\ i}^{i^{\prime }}$ which, in
its turn, allows us to compute $N_{i^{\prime }}^{a^{\prime }}$(\ref{defcncl}%
) and all coefficients of the metric $\mathbf{g}$ (\ref{gpsm}) and vierbein
transform (\ref{dft}). We shall omit for simplicity the left labe $L$ if
that will not result in a confusion for some special construction, but we
shall always keep in mind that a generating function (not brokening the
general diffeomorphysm symmetry) can be introduces if it would be necessary
for an effective Lagrange modelling of (pseudo) Riemannian constructions.
\footnote{%
We emphasize, that any object labeled with a left symbol $L$ can be defined
geometrically for any generating function $L.$ This mean that the
constructions are diffeomorphic invariant and do not depend in explicit form
on $L;$ like in ADM formalism, there is not dependence of geometric
constructins on the type of slac and shift functions. In both cases, such
prescribed functions are useful for establishing cetain coordinate and frame
forumulas, with respective 2+2 and 3+1 splitting, but the definition of
almost K\"{a}hler and Ashterkar connections is independent on the explicit
type of fibration.}

\subsubsection{Canonical Lagrange nonlinear connections in GR}

One holds some important results on analogous geometric mechanical modelling
of gravitational interactions (see details in Refs. \cite%
{ma,vsgg,vrfg,vqg1,vqg2,vqg3,vqg4,avqg5}):

The Euler--Lagrange equations
\begin{equation*}
\frac{d}{d\tau }\left( \frac{\partial L}{\partial y^{i}}\right) -\frac{%
\partial L}{\partial x^{i}}=0
\end{equation*}%
are equivalent to the ''nonlinear'' geodesic equations%
\begin{equation*}
\frac{dy^{a}}{d\tau }+2G^{a}(x^{k},y^{b})=0,
\end{equation*}%
where $G^{a}$ is computed following formula (\ref{clssp}). The solutions of
these equation, parametrized as $u^{\alpha }(\tau )=\left( x^{i}(\tau
),y^{a}(\tau )\right) $ for a real parameter $0\leq \tau \leq \tau _{0},$
when $y^{a}(\tau )=dx^{i}(\tau )/d\tau ,$ define the paths of the canonical
semispray
\begin{equation}
S=y^{i}\frac{\partial }{\partial x^{i}}-2G^{a}\frac{\partial }{\partial y^{a}%
}  \label{sspray}
\end{equation}%
covering the region $U\subset \mathbf{V}$ with a prescribed regular
effective Lagrangian.

Having chosen on $\mathbf{V}$ an effective Lagrange structure $L,$ we
associate to any (pseudo) Riemannian metric $\mathbf{g}$ (\ref{gpsm}),
equivalently $^{L}\mathbf{g}$ (\ref{lfsm}), a set of important geometric
objects as in usual Lagrange mechanics (this is a formal analogy because we
do not work on a tangent bundle):

A nonlinear connection (N--connection) $\mathbf{N}$ is induced as a Whitney
sum (nonholonomic distribution)
\begin{equation}
T\mathbf{V}=h\mathbf{V}\oplus v\mathbf{V},  \label{whitney}
\end{equation}%
splitting globally the tangent bundle $T\mathbf{V}$ into respective h-- and
v--subspac\-es, $h\mathbf{V}$ and $v\mathbf{V},$ given locally, in canonical
form defined by a regular $L,$ by a set of coefficients $^{L}N_{i}^{a}$ (\ref%
{clnc}), equivalently $N_{i^{\prime }}^{a^{\prime }}$ (\ref{defcncl}), \ as
\begin{equation*}
\ ^{L}\mathbf{N=}\ ^{L}N_{i}^{a}dx^{i}\otimes \frac{\partial }{\partial y^{a}%
}=\ N_{i^{\prime }}^{a^{\prime }}e^{i^{\prime }}\otimes e_{a^{\prime }}.
\end{equation*}%
There is a proof in Ref. \cite{ma} that for any vector bundle over a
paracompact manifold there exists nonlinear connections. In this work, we
shall consider only (pseudo) Riemannian manifolds admitting a local fibred
structure with a splitting $\ ^{L}\mathbf{N}$ defined by coefficients of
type (\ref{defcncl}).

For any N--connection structure $\mathbf{N}=\{N_{j}^{a}\},$ we can define $\
$on $\mathbf{V}$ a frame structure (with coefficients depending linearly on $%
N_{j}^{a})$ denoted $\mathbf{e}_{\nu }=(\mathbf{e}_{i},e_{a}),$ where
\begin{equation}
\mathbf{e}_{i}=\frac{\partial }{\partial x^{i}}-N_{i}^{a}(u)\frac{\partial }{%
\partial y^{a}}\mbox{ and
}e_{a}=\frac{\partial }{\partial y^{a}},  \label{dder}
\end{equation}%
and the dual frame (coframe) structure is $\mathbf{e}^{\mu }=(e^{i},\mathbf{e%
}^{a}),$ where
\begin{equation}
e^{i}=dx^{i}\mbox{ and }\mathbf{e}^{a}=dy^{a}+N_{i}^{a}(u)dx^{i},
\label{ddif}
\end{equation}%
satisfying nontrivial nonholonomy relations
\begin{equation}
\lbrack \mathbf{e}_{\alpha },\mathbf{e}_{\beta }]=\mathbf{e}_{\alpha }%
\mathbf{e}_{\beta }-\mathbf{e}_{\beta }\mathbf{e}_{\alpha }=W_{\alpha \beta
}^{\gamma }\mathbf{e}_{\gamma },  \label{anhrel}
\end{equation}%
with (antisymmetric) anholonomy coefficients $W_{ia}^{b}=\partial
_{a}N_{i}^{b}$ and $W_{ji}^{a}=\Omega _{ij}^{a}.$ We omitted the label $L$
for the above frame coefficients formulas in order to emphasize that such
geometric objects can be defined for any formal splitting (\ref{whitney})
induced, or not, by an effective Lagrangian.

Any effective regular Lagrangian $L$ prescribed on a spacetime manifold $%
\mathbf{V}$ states a canonical N--connection, $\ ^{L}\mathbf{N}=\{\ ^{L}N_{\
\underline{i}}^{a^{\prime }}(u)\}$ (\ref{clnc}), and frame, $\mathbf{e}_{\nu
}=(\mathbf{e}_{i},e_{a^{\prime }})$ and $\mathbf{e}^{\mu }=(e^{i},\mathbf{e}%
^{a^{\prime }}),$ structures. In out further considerations, we shall put
the label $L$ only \ in the cases when it is important to \ emphasize
certain geometric structures/objects are induced by a regular $L$ and omit
left labels if the formulas will hold true for more general classes of
nonholonomic distributions.

One perform N--adapted geometric constructions by defining the coefficients
of geometric objects and respective equations with respect to noholonomic
frames of type (\ref{dder}) and (\ref{ddif}). The N--adapted tensors,
vectors, forms etc are called respectively distinguished tensors (by a
N--connection structure) etc, in brief, d--tensors, d--vectors, d--forms
etc. For instance, a vector field $\mathbf{X}\in T\mathbf{V} $ is expressed
\begin{equation*}
\mathbf{X}=(hX,\ vX),\mbox{ \ or \ }\mathbf{X}=X^{\alpha }\mathbf{e}_{\alpha
}=X^{i}\mathbf{e}_{i}+X^{a}e_{a},
\end{equation*}%
where $hX=X^{i}\mathbf{e}_{i}$ and $vX=X^{a}e_{a}$ state, respectively, the
adapted to the N--connection structure horizontal (h) and vertical (v)
components of the vector.

As a particular case of Lagrange modelling, we may consider a Finsler
modelling with $L=F^{2}(x,y),$ where the effective Finsler metric $F$
(fundamental function) is a differentiable of class $C^{\infty }$ in any
point $(x,y)$ with $y\neq 0$ and is continuous in any points $(x,0);$ $%
F(x,y)>0$ if $y\neq 0;$ it is satisfied the homogeneity condition $F(x,\beta
y)=|\beta |F(x,y)$ for any nonzero $\beta \in \mathbb{R},$ and the Hessian (%
\ref{elm}) computed for $L=F^{2}$ is positive definite. In such cases, we
argue that a (pseudo) Riemannian space with metric $\mathbf{g}$ (\ref{gpsm})
is modelled by an effective Finsler geometry and, inversely, a Finsler
geometry is modelled on a (pseudo) Riemannian space. Similar constructions
were performed, for instance, in Ref. \cite{ma} for Lagrange and Finsler
spaces defined on tangent bundles. In our works, we follow an approach when
such geometries are modelled on (pseudo) Riemannian and Riemann--Cartan
spaces endowed with nonholonomic distributions, see a review of results in
Ref. \cite{vrfg}.

A (pseudo) Riemannian manifold $\mathbf{V}$ is nonholonomic
(N--an\-holonomic) if it is provided with a nonholonomic distribution on $T%
\mathbf{V}$ (N--connection structure $\mathbf{N).}$ Any (pseudo) Riemannian
space can be transformed into a N--anho\-lo\-nomic manifold $\mathbf{V}$
modelling an effective Lagrange (or Finsler) geometry by prescribing a
generating Lagrange (or Finsler) function $L(x,y)$ (or $F(x,y)).$ For
simplicity, in this work we shall use only Lagrange structures considering
the Finsler ones to consist a particular (homogeneous) case.

\subsubsection{Canonical almost complex/ symplectic structures}

Let $\mathbf{e}_{\alpha ^{\prime }}=(\mathbf{e}_{i},e_{b^{\prime }})$ and $%
\mathbf{e}^{\alpha ^{\prime }}=(e^{i},\ \mathbf{e}^{b^{\prime }})$ be
defined respectively by (\ref{dder}) and (\ref{ddif}) for the canonical
N--connection $\ ^{L}\mathbf{N}$ (\ref{clnc}) stated by a metric structure $%
\mathbf{g}=$ $~^{L}\mathbf{g}$ (\ref{lfsm}) on $\mathbf{V}.$ We introduce a
linear operator $\mathbf{J}$ acting on vectors on $\mathbf{V}$ following
formulas
\begin{equation*}
\mathbf{J}(\mathbf{e}_{i})=-e_{2+i}\mbox{\ and \ }\mathbf{J}(e_{2+i})=%
\mathbf{e}_{i},
\end{equation*}%
where and $\mathbf{J\circ J=-I}$ for $\mathbf{I}$ being the unity matrix,
and construct a tensor field on $\mathbf{V},$%
\begin{eqnarray}
\mathbf{J} &=&\mathbf{J}_{\ \beta }^{\alpha }\ e_{\alpha }\otimes e^{\beta }=%
\mathbf{J}_{\ \underline{\beta }}^{\underline{\alpha }}\ \frac{\partial }{%
\partial u^{\underline{\alpha }}}\otimes du^{\underline{\beta }}
\label{acstr} \\
&=&\mathbf{J}_{\ \beta ^{\prime }}^{\alpha ^{\prime }}\ \mathbf{e}_{\alpha
^{\prime }}\otimes \mathbf{e}^{\beta ^{\prime }}=\mathbf{-}e_{2+i}\otimes
e^{i}+\mathbf{e}_{i}\otimes \ \mathbf{e}^{2+i}  \notag \\
&=&-\frac{\partial }{\partial y^{i}}\otimes dx^{i}+\left( \frac{\partial }{%
\partial x^{i}}-\ ^{L}N_{i}^{2+j}\frac{\partial }{\partial y^{j}}\right)
\otimes \left( dy^{i}+\ ^{L}N_{k}^{2+i}dx^{k}\right) ,  \notag
\end{eqnarray}%
defining globally an almost complex structure on\ $\mathbf{V}$ completely
determined by a fixed $L(x,y).$ Using vielbeins $\mathbf{e}_{\ \underline{%
\alpha }}^{\alpha }$ and their duals $\mathbf{e}_{\alpha \ }^{\ \underline{%
\alpha }}$, defined by $e_{\ \underline{i}}^{i},e_{\ \underline{i}}^{a}$ and
$e_{\ \underline{a}}^{a}$ solving (\ref{algeq1})--(\ref{algeq3}), we can
compute the coefficients of tensor $\mathbf{J}$ with respect to any local
basis $e_{\alpha }$ and $e^{\alpha }$ on $\mathbf{V},$ $\mathbf{J}_{\ \beta
}^{\alpha }=\mathbf{e}_{\ \underline{\alpha }}^{\alpha }\mathbf{J}_{\
\underline{\beta }}^{\underline{\alpha }}\mathbf{e}_{\beta \ }^{\ \underline{%
\beta }}.$ In general, we can define an almost complex structure $\mathbf{J}$
for an arbitrary N--connection $\mathbf{N}$ by using N--adapted bases (\ref%
{dder}) and (\ref{ddif}), not obligatory induced by an effective Lagrange
function.

The Neijenhuis tensor field for any almost complex structure $\mathbf{J}$
defined by a N--connection (equivalently, the curvature of N--connecti\-on)
is
\begin{equation}
\ ^{\mathbf{J}}\mathbf{\Omega (X,Y)\doteqdot -[X,Y]+[JX,JY]-J[JX,Y]-J[X,JY],}
\label{neijt}
\end{equation}%
for any d--vectors $\mathbf{X}$ and $\mathbf{Y.}$ With respect to N--adapted
bases (\ref{dder}) and (\ref{ddif}), a subset of the coefficients of the
Neijenhuis tensor defines the N--connection curvature, see details in Ref. %
\cite{ma},
\begin{equation}
\Omega _{ij}^{a}=\frac{\partial N_{i}^{a}}{\partial x^{j}}-\frac{\partial
N_{j}^{a}}{\partial x^{i}}+N_{i}^{b}\frac{\partial N_{j}^{a}}{\partial y^{b}}%
-N_{j}^{b}\frac{\partial N_{i}^{a}}{\partial y^{b}}.  \label{nccurv}
\end{equation}%
A N--anholonomic manifold $\mathbf{V}$ is integrable if $\Omega _{ij}^{a}=0.$
We get a complex structure if and only if both the h-- and v--distributions
are integrable, i.e. if and only if $\Omega _{ij}^{a}=0$ and $\frac{\partial
N_{j}^{a}}{\partial y^{i}}-\frac{\partial N_{i}^{a}}{\partial y^{j}}=0.$

One calls an almost symplectic structure on a manifold $V$ a nondegenerate
2--form
\begin{equation*}
\theta =\frac{1}{2}\theta _{\alpha \beta }(u)e^{\alpha }\wedge e^{\beta }.
\end{equation*}%
For any $\theta $ on $V,$ there is a unique N--connection $\mathbf{N}%
=\{N_{i}^{a}\}$ (\ref{whitney}) satisfying the conditions:%
\begin{equation}
\theta =(h\mathbf{X},v\mathbf{Y})=0\mbox{ and }\theta =h\theta +v\theta ,
\label{aux02}
\end{equation}%
for any $\mathbf{X}=h\mathbf{X}+v\mathbf{X,}$ $\mathbf{Y}=h\mathbf{Y}+v%
\mathbf{Y},$ where $h\theta (\mathbf{X,Y})\doteqdot \theta (h\mathbf{X,}h%
\mathbf{Y})$ and $v\theta (\mathbf{X,Y})\doteqdot \theta (v\mathbf{X,}v%
\mathbf{Y}).$

For $\mathbf{X=e}_{\alpha }=(\mathbf{e}_{i},e_{a})$ and $\mathbf{Y=e}_{\beta
}=(\mathbf{e}_{l},e_{b}),$ where $\mathbf{e}_{\alpha }$ is a N--adapted
basis\ of type (\ref{dder}), we write the first equation in (\ref{aux02}) in
the form%
\begin{equation*}
\theta =\theta (\mathbf{e}_{i},e_{a})=\theta (\frac{\partial }{\partial x^{i}%
},\frac{\partial }{\partial y^{a}})-N_{i}^{b}\theta (\frac{\partial }{%
\partial y^{b}},\frac{\partial }{\partial y^{a}})=0.
\end{equation*}%
We can solve this system of equations in a unique form and define $N_{i}^{b}$
if $rank|\theta (\frac{\partial }{\partial y^{b}},\frac{\partial }{\partial
y^{a}})|=2.$ Denoting locally
\begin{equation}
\mathbf{\theta }=\frac{1}{2}\theta _{ij}(u)e^{i}\wedge e^{j}+\frac{1}{2}%
\theta _{ab}(u)\mathbf{e}^{a}\wedge \mathbf{e}^{b},  \label{aux03}
\end{equation}%
where the first term is for $h\theta $ and the second term is $v\theta ,$ we
get the second formula in (\ref{aux02}).

An almost Hermitian model of a (pseudo) Riemannian spa\-ce $\mathbf{V}$
equipped with a N--connection structure $\mathbf{N}$ is defined by a triple $%
\mathbf{H}^{2+2}=(\mathbf{V},\theta ,\mathbf{J}),$ where $\mathbf{\theta
(X,Y)}\doteqdot \mathbf{g}\left( \mathbf{JX,Y}\right) $ for any $\mathbf{g}$
(\ref{gpsm}). A space $\mathbf{H}^{2+2}$ is almost K\"{a}hler, denoted $%
\mathbf{K}^{2+2},$ if and only if $d\mathbf{\theta }=0.$

For $\mathbf{g}=\ $ $^{L}\mathbf{g}$ (\ref{lfsm}) and structures $\ ^{L}%
\mathbf{N}$ (\ref{clnc}) and $\mathbf{J}$ canonically defined by $L,$ we
define $\ ^{L}\mathbf{\theta (X,Y)}\doteqdot \ ^{L}\mathbf{g}\left( \mathbf{%
JX,Y}\right) $ for any d--vectors $\mathbf{X}$ and $\mathbf{Y.}$ In local
N--adapted form form, we have
\begin{eqnarray}
\ ^{L}\mathbf{\theta } &=&\frac{1}{2}\ ^{L}\theta _{\alpha \beta
}(u)e^{\alpha }\wedge e^{\beta }=\frac{1}{2}\ ^{L}\theta _{\underline{\alpha
}\underline{\beta }}(u)du^{\underline{\alpha }}\wedge du^{\underline{\beta }}
\label{asymstr} \\
&=&\ ^{L}g_{ij}(x,y)e^{2+i}\wedge dx^{j}=\ ^{L}g_{ij}(x,y)(dy^{2+i}+\
^{L}N_{k}^{2+i}dx^{k})\wedge dx^{j}.  \notag
\end{eqnarray}%
Let us consider the form $\ ^{L}\omega =\frac{1}{2}\frac{\partial L}{%
\partial y^{i}}dx^{i}.$ A straightforward computation shows that $\ ^{L}%
\mathbf{\theta }=d\ ^{L}\omega ,$ which means that $d\ ^{L}\mathbf{\theta }%
=dd\ ^{L}\omega =0,$ i.e. the canonical effective Lagrange structures $%
\mathbf{g}=\ ^{L}\mathbf{g},\ \ ^{L}\mathbf{N}$ and $\mathbf{J}$ induce an
almost K\"{a}hler geometry. We can express the 2--form (\ref{asymstr}) as
\begin{equation*}
\mathbf{\theta }=\ ^{L}\mathbf{\theta }=\frac{1}{2}\ ^{L}\theta
_{ij}(u)e^{i}\wedge e^{j}+\frac{1}{2}\ ^{L}\theta _{ab}(u)\mathbf{e}%
^{a}\wedge \mathbf{e}^{b},
\end{equation*}%
see (\ref{aux03}), where the coefficients $\ ^{L}\theta _{ab}=\ ^{L}\theta
_{2+i\ 2+j}$ are equal respectively to the coefficients $\ ^{L}\theta _{ij}.$
It should be noted that for a general 2--form $\theta $ constructed for any
metric $\mathbf{g}$ and almost complex $\mathbf{J}$\textbf{\ }structures on $%
V$ one holds $d\theta \neq 0.$ But for any $2+2$ splitting induced by an
effective Lagrange generating function, we have $d\ ^{L}\mathbf{\theta }=0.$
We have also $d\ \mathbf{\theta }=0$ for any set of 2--form coefficients $%
\mathbf{\theta }_{\alpha ^{\prime }\beta ^{\prime }}e_{\ \alpha }^{\alpha
^{\prime }}e_{\ \beta }^{\beta ^{\prime }}=~^{L}\mathbf{g}_{\alpha ^{\prime
}\beta ^{\prime }},$ constructed by using formulas (\ref{algeq1})--(\ref%
{algeq3}).

We conclude that having chosen a generating function $L(x,y)$ on a (pseudo)
Riemannian spacetime $\mathbf{V},$ we can model this spacetime equivalently
as an almost K\"{a}hler manifold (more exactly, for corresponding generating
functions, as an almost K\"{a}hler--Lagrange, or K\"{a}hler--Finsler,
nonholonomic manifold).

\subsubsection{Equivalent metric compatible linear connections}

A distinguished connection (in brief, d--connection) on a spacetime $\mathbf{%
V}$,
\begin{equation*}
\mathbf{D}=(hD;vD)=\{\mathbf{\Gamma }_{\beta \gamma }^{\alpha
}=(L_{jk}^{i},\ ^{v}L_{bk}^{a};C_{jc}^{i},\ ^{v}C_{bc}^{a})\},
\end{equation*}%
is a linear connection which preserves under parallel transports the
distribution (\ref{whitney}). In explicit form, the coefficients\ $\mathbf{%
\Gamma }_{\beta \gamma }^{\alpha }$ are computed with respect to a
N--adapted basis (\ref{dder}) and (\ref{ddif}). A d--connection $\mathbf{D}$%
\ is metric compatible with a d--metric $\mathbf{g}$ if $\mathbf{D}_{\mathbf{%
X}}\mathbf{g}=0$ for any d--vector field $\mathbf{X.}$

If an almost symplectic structure $\theta $ is considered on a
N--anholonomic manifold, an almost symplectic d--connection $\ _{\theta }%
\mathbf{D}$ on $\mathbf{V}$ is defined by the conditions that it is
N--adapted, i.e. it is a d--connection, and $\ _{\theta }\mathbf{D}_{\mathbf{%
X}}\theta =0,$ for any d--vector $\mathbf{X.}$ From the set of metric and/or
almost symplectic compatible d--connecti\-ons on a (pseudo) Riemannian
manifold $\mathbf{V},$ we can select those which are completely defined by a
metric $\mathbf{g}=\ $ $^{L}\mathbf{g}$ (\ref{lfsm}) and an effective
Lagrange structure $L(x,y):$

There is a unique normal d--connection
\begin{eqnarray}
\ \widehat{\mathbf{D}} &=&\left\{ h\widehat{D}=(\widehat{D}_{k},^{v}\widehat{%
D}_{k}=\widehat{D}_{k});v\widehat{D}=(\widehat{D}_{c},\ ^{v}\widehat{D}_{c}=%
\widehat{D}_{c})\right\}  \label{ndc} \\
&=&\{\widehat{\mathbf{\Gamma }}_{\beta \gamma }^{\alpha }=(\widehat{L}%
_{jk}^{i},\ ^{v}\widehat{L}_{2+j\ 2+k}^{2+i}=\widehat{L}_{jk}^{i};\ \widehat{%
C}_{jc}^{i}=\ ^{v}\widehat{C}_{2+j\ c}^{2+i},\ ^{v}\widehat{C}_{bc}^{a}=%
\widehat{C}_{bc}^{a})\},  \notag
\end{eqnarray}%
which is metric compatible,
\begin{equation*}
\widehat{D}_{k}\ ^{L}g_{ij}=0\mbox{ and }\widehat{D}_{c}\ ^{L}g_{ij}=0,
\end{equation*}%
and completely defined by a couple of h-- and v--components $\ \widehat{%
\mathbf{D}}_{\alpha }=(\widehat{D}_{k},\widehat{D}_{c}),$ with N--adapted
coefficients $\widehat{\mathbf{\Gamma }}_{\beta \gamma }^{\alpha }=(\widehat{%
L}_{jk}^{i},\ ^{v}\widehat{C}_{bc}^{a}),$ where
\begin{eqnarray}
\widehat{L}_{jk}^{i} &=&\frac{1}{2}\ ^{L}g^{ih}\left( \mathbf{e}_{k}\
^{L}g_{jh}+\mathbf{e}_{j}\ ^{L}g_{hk}-\mathbf{e}_{h}\ ^{L}g_{jk}\right) ,
\label{cdcc} \\
\widehat{C}_{jk}^{i} &=&\frac{1}{2}\ ^{L}g^{ih}\left( \frac{\partial \
^{L}g_{jh}}{\partial y^{k}}+\frac{\partial \ ^{L}g_{hk}}{\partial y^{j}}-%
\frac{\partial \ ^{L}g_{jk}}{\partial y^{h}}\right) .  \notag
\end{eqnarray}%
In general, we can ''foget'' about label $L$ and work with arbitrary $%
\mathbf{g}_{\alpha ^{\prime }\beta ^{\prime }}$ and $\widehat{\mathbf{\Gamma
}}_{\beta ^{\prime }\gamma ^{\prime }}^{\alpha ^{\prime }}$ with the
coefficients recomputed by frame transforms (\ref{algeq1})--(\ref{algeq3}).

Introducing the normal d--connection 1--form%
\begin{equation*}
\widehat{\mathbf{\Gamma }}_{j}^{i}=\widehat{L}_{jk}^{i}e^{k}+\widehat{C}%
_{jk}^{i}\mathbf{e}^{k},
\end{equation*}%
we prove that the Cartan structure equations are satisfied,%
\begin{equation}
de^{k}-e^{j}\wedge \widehat{\mathbf{\Gamma }}_{j}^{k}=-\widehat{\mathcal{T}}%
^{i},\ d\mathbf{e}^{k}-\mathbf{e}^{j}\wedge \widehat{\mathbf{\Gamma }}%
_{j}^{k}=-\ ^{v}\widehat{\mathcal{T}}^{i},  \label{cart1}
\end{equation}%
and
\begin{equation}
d\widehat{\mathbf{\Gamma }}_{j}^{i}-\widehat{\mathbf{\Gamma }}_{j}^{h}\wedge
\widehat{\mathbf{\Gamma }}_{h}^{i}=-\widehat{\mathcal{R}}_{\ j}^{i}.
\label{cart2}
\end{equation}%
The h-- and v--components of the torsion 2--form
\begin{equation*}
\widehat{\mathcal{T}}^{\alpha }=\left( \widehat{\mathcal{T}}^{i},\ ^{v}%
\widehat{\mathcal{T}}^{i}\right) =\widehat{\mathbf{T}}_{\ \tau \beta
}^{\alpha }\ \mathbf{e}^{\tau }\wedge \mathbf{e}^{\beta }
\end{equation*}
from (\ref{cart1}) is computed with components
\begin{equation}
\widehat{\mathcal{T}}^{i}=\widehat{C}_{jk}^{i}e^{j}\wedge \mathbf{e}^{k},\
^{v}\widehat{\mathcal{T}}^{i}=\frac{1}{2}\ ^{L}\Omega _{kj}^{i}e^{k}\wedge
e^{j}+(\frac{\partial \ \ ^{L}N_{k}^{i}}{\partial y^{j}}-\widehat{L}_{\
kj}^{i})e^{k}\wedge \mathbf{e}^{j},  \label{tform}
\end{equation}%
where $\ ^{L}\Omega _{kj}^{i}$ are coefficients of the curvature of the
canonical N--connection $\check{N}_{k}^{i}$ defined by formulas similar to (%
\ref{nccurv}). The formulas (\ref{tform}) parametrize the h-- and
v--components of torsion $\widehat{\mathbf{T}}_{\beta \gamma }^{\alpha }$ in
the form
\begin{equation}
\widehat{T}_{jk}^{i}=0,\widehat{T}_{jc}^{i}=\widehat{C}_{\ jc}^{i},\widehat{T%
}_{ij}^{a}=\ ^{L}\Omega _{ij}^{a},\widehat{T}_{ib}^{a}=e_{b}\left( \
^{L}N_{i}^{a}\right) -\widehat{L}_{\ bi}^{a},\widehat{T}_{bc}^{a}=0.
\label{cdtors}
\end{equation}%
It should be noted that $\widehat{\mathbf{T}}$ vanishes on h- and
v--subspaces, i.e. $\widehat{T}_{jk}^{i}=0$ and $\widehat{T}_{bc}^{a}=0,$
but certain nontrivial h--v--components induced by the nonholonomic
structure are defined canonically by $\mathbf{g}=\ ^{L}\mathbf{g}$ (\ref%
{lfsm}) and $L.$

We compute also the curvature 2--form from (\ref{cart2}),%
\begin{eqnarray}
\widehat{\mathcal{R}}_{\ \gamma }^{\tau } &=&\widehat{\mathbf{R}}_{\ \gamma
\alpha \beta }^{\tau }\ \mathbf{e}^{\alpha }\wedge \ \mathbf{e}^{\beta }
\label{cform} \\
&=&\frac{1}{2}\widehat{R}_{\ jkh}^{i}e^{k}\wedge e^{h}+\widehat{P}_{\
jka}^{i}e^{k}\wedge \mathbf{e}^{a}+\frac{1}{2}\ \widehat{S}_{\ jcd}^{i}%
\mathbf{e}^{c}\wedge \mathbf{e}^{d},  \notag
\end{eqnarray}%
where the nontrivial N--adapted coefficients of curvature $\widehat{\mathbf{R%
}}_{\ \beta \gamma \tau }^{\alpha }$ of $\widehat{\mathbf{D}}$ are
\begin{eqnarray}
\widehat{R}_{\ hjk}^{i} &=&\mathbf{e}_{k}\widehat{L}_{\ hj}^{i}-\mathbf{e}%
_{j}\widehat{L}_{\ hk}^{i}+\widehat{L}_{\ hj}^{m}\widehat{L}_{\ mk}^{i}-%
\widehat{L}_{\ hk}^{m}\widehat{L}_{\ mj}^{i}-\widehat{C}_{\ ha}^{i}\
^{L}\Omega _{\ kj}^{a}  \label{cdcurv} \\
\widehat{P}_{\ jka}^{i} &=&e_{a}\widehat{L}_{\ jk}^{i}-\widehat{\mathbf{D}}%
_{k}\widehat{C}_{\ ja}^{i},  \notag \\
\widehat{S}_{\ bcd}^{a} &=&e_{d}\widehat{C}_{\ bc}^{a}-e_{c}\widehat{C}_{\
bd}^{a}+\widehat{C}_{\ bc}^{e}\widehat{C}_{\ ed}^{a}-\widehat{C}_{\ bd}^{e}%
\widehat{C}_{\ ec}^{a}.  \notag
\end{eqnarray}%
Contracting the first and forth indices $\widehat{\mathbf{\mathbf{R}}}%
\mathbf{_{\ \beta \gamma }=}\widehat{\mathbf{\mathbf{R}}}\mathbf{_{\ \beta
\gamma \alpha }^{\alpha }}$, we get the N--adapted coefficients for the
Ricci tensor%
\begin{equation}
\widehat{\mathbf{\mathbf{R}}}\mathbf{_{\beta \gamma }=}\left( \widehat{R}%
_{ij},\widehat{R}_{ia},\widehat{R}_{ai},\widehat{R}_{ab}\right) .
\label{dricci}
\end{equation}%
The scalar curvature $\ ^{L}R=\widehat{R}$ of $\widehat{\mathbf{D}}$ is
\begin{equation}
\ ^{L}R=\ ^{L}\mathbf{g}^{\beta \gamma }\widehat{\mathbf{\mathbf{R}}}\mathbf{%
_{\beta \gamma }=\ \mathbf{g}^{\beta ^{\prime }\gamma ^{\prime }}\widehat{%
\mathbf{\mathbf{R}}}\mathbf{_{\beta ^{\prime }\gamma ^{\prime }}}.}
\label{scalc}
\end{equation}

The normal d--connection $\widehat{\mathbf{D}}$ (\ref{ndc}) defines a
canonical almost symplectic d--connection, $\widehat{\mathbf{D}}\equiv \
_{\theta }\widehat{\mathbf{D}},$ which is N--adapted to the effective
Lagrange and, related, almost symplectic structures, i.e. it preserves under
parallelism the splitting (\ref{whitney}), $_{\theta }\widehat{\mathbf{D}}_{%
\mathbf{X}}\ ^{L}\mathbf{\theta =}_{\theta }\widehat{\mathbf{D}}_{\mathbf{X}%
}\ \mathbf{\theta =}0$ and its torsion is constrained to satisfy the
conditions $\widehat{T}_{jk}^{i}=\widehat{T}_{bc}^{a}=0.$ In the canonical
approach to the general relativity theory, one works with the Levi Civita
connection $\bigtriangledown =\{\ _{\shortmid }\Gamma _{\beta \gamma
}^{\alpha }\}$ which is uniquely derived following the conditions $~\
_{\shortmid }\mathcal{T}=0$ and $\bigtriangledown \mathbf{g}=0.$ This is a
linear connection but not a d--connection because $\bigtriangledown $ does
not preserve (\ref{whitney}) under parallelism. Both linear connections $%
\bigtriangledown $ and $\widehat{\mathbf{D}}\equiv \ _{\theta }\widehat{%
\mathbf{D}}$ are uniquely defined in metric compatible forms by the same
metric structure $\mathbf{g}$ (\ref{gpsm}). The second one contains
nontrivial d--torsion components $\widehat{\mathbf{T}}_{\beta \gamma
}^{\alpha }$ (\ref{cdtors}), induced effectively by an equivalent Lagrange
metric $\mathbf{g}=\ ^{L}\mathbf{g}$ (\ref{lfsm}) and adapted both to the
N--connection $\ ^{L}\mathbf{N,}$ see (\ref{clnc}) and (\ref{whitney}), and
almost symplectic $\ ^{L}\mathbf{\theta }$ (\ref{asymstr}) structures $L.$

Any geometric construction for the normal d--connection $\widehat{\mathbf{D}}
$ can be re--defined by the Levi Civita connection, and inversely, using the
formula
\begin{equation}
\ _{\shortmid }\Gamma _{\ \alpha \beta }^{\gamma }(\mathbf{g})=\widehat{%
\mathbf{\Gamma }}_{\ \alpha \beta }^{\gamma }(\mathbf{g})+\ _{\shortmid
}Z_{\ \alpha \beta }^{\gamma }(\mathbf{g}),  \label{cdeft}
\end{equation}%
where the both connections $\ _{\shortmid }\Gamma _{\ \alpha \beta }^{\gamma
}(\mathbf{g})$ and $\widehat{\mathbf{\Gamma }}_{\ \alpha \beta }^{\gamma }(%
\mathbf{g})$ and the distorsion tensor $\ _{\shortmid }Z_{\ \alpha \beta
}^{\gamma }(\mathbf{g})$ with N--adapted coefficients (for the normal
d--connection $\ _{\shortmid }Z_{\ \alpha \beta }^{\gamma }(\mathbf{g})$ is
proportional to $\widehat{\mathbf{T}}_{\beta \gamma }^{\alpha }(\mathbf{g})$
(\ref{cdtors})). In this work, we emphasize if it is necessary the
functional dependence of certain geometric objects on a d--metric $(\mathbf{g%
})$, for tensors and connections completely defined by the metric structure.%
\footnote{%
see formulas (\ref{defashtd}), and below, in Appendix, on similar
deformation properties of fundamental geometric objects}

If we work with nonholonomic constraints on the dynamics/ geometry of
gravity fields in DQ, it is more convenient to use a N--adapted and/or
almost symplectic approach. For other purposes, it is preferred to use only
the Levi--Civita connection, or a 3+1 formalism, for instance, in ADM
formulation of gravity. Introducing the distorsion relation (\ref{cdeft})
into respective formulas (\ref{cdtors}), (\ref{cdcurv}) and (\ref{dricci})
written for $\widehat{\mathbf{\Gamma }}_{\ \alpha \beta }^{\gamma },$ we get
deformations
\begin{eqnarray}
\ _{\shortmid }T_{\ \beta \gamma }^{\alpha }(\mathbf{g}) &=&\widehat{\mathbf{%
T}}_{\ \beta \gamma }^{\alpha }(\mathbf{g})+\ _{\shortmid }Z_{\ \alpha \beta
}^{\gamma }(\mathbf{g})=0,  \label{aux01} \\
\ _{\shortmid }R_{\ \beta \gamma \delta }^{\alpha }(\mathbf{g}) &=&\widehat{%
\mathbf{R}}_{\ \beta \gamma \delta }^{\alpha }+\ _{\shortmid }\widehat{%
\mathbf{Z}}_{\ \beta \gamma \delta }^{\alpha }(\mathbf{g}),\ _{\shortmid
}R_{\ \beta \gamma }(\mathbf{g})=\widehat{\mathbf{R}}_{\ \beta \gamma }+\
_{\shortmid }\widehat{\mathbf{Z}}_{\ \beta \gamma }(\mathbf{g}),  \notag
\end{eqnarray}%
see Refs. \cite{vrfg,vsgg} and Appendix to this work for explicit formulas
for distorisons of the torsion, curvature, Ricci tensors, i.e. $\
_{\shortmid }^{T}Z_{\ \alpha \beta }^{\gamma }(\mathbf{g}),$ $\ _{\shortmid }%
\widehat{\mathbf{Z}}_{\ \beta \gamma \delta }^{\alpha }(\mathbf{g})$ and $\
_{\shortmid }\widehat{\mathbf{Z}}_{\ \beta \gamma }(\mathbf{g}),$ which are
completely defined by a metric structure $\mathbf{g}=\ ^{L}\mathbf{g}$ with
a nonholonomic 2+2 splitting induced by a prescribed regular $L.$

\subsection{An almost symplectic formulation of GR}

\label{ssakgr}Having chosen a canonical almost symplectic d--connection, we
compute the Ricci d--tensor $\widehat{\mathbf{R}}_{\ \beta \gamma }$ (\ref%
{dricci}) and the scalar curvature $\ ^{L}R$ $\ $(\ref{scalc})). Then, we
can postulate in a straightforward form the filed equations
\begin{equation}
\widehat{\mathbf{R}}_{\ \beta }^{\underline{\alpha }}-\frac{1}{2}(\
^{L}R+\lambda )\mathbf{e}_{\ \beta }^{\underline{\alpha }}=8\pi G\mathbf{T}%
_{\ \beta }^{\underline{\alpha }},  \label{deinsteq}
\end{equation}%
where $\widehat{\mathbf{R}}_{\ \ \beta }^{\underline{\alpha }}=\mathbf{e}_{\
\gamma }^{\underline{\alpha }}$ $\widehat{\mathbf{R}}_{\ \ \beta }^{\ \gamma
},$ $\mathbf{T}_{\ \beta }^{\underline{\alpha }}$ is the effective
energy--momentum tensor, $\lambda $ is the cosmological constant, $G$ is the
Newton constant in the units when the light velocity $c=1,$ and the
coefficients $\mathbf{e}_{\ \beta }^{\underline{\alpha }}$ of vierbein
decomposition $\mathbf{e}_{\ \beta }=\mathbf{e}_{\ \beta }^{\underline{%
\alpha }}\partial /\partial u^{\underline{\alpha }}$ are defined by the
N--coefficients of the N--elongated operator of partial derivation, see (\ref%
{dder}).

In order to formulate a variational N--adapted formalism for equations (\ref%
{deinsteq}), for the gravitational $\left( \mathbf{e,}\widehat{\mathbf{%
\Gamma }}\right) $ and matter $\mathbf{\phi }$ fields, we consider the
effective action
\begin{equation}
\mathcal{S}[\mathbf{e,\widehat{\mathbf{\Gamma }},\phi }]=\ ^{gr}\mathcal{S}[%
\mathbf{e,}\widehat{\mathbf{\Gamma }}]+\ ^{matter}\mathcal{S}[\mathbf{e,%
\widehat{\mathbf{\Gamma }},\phi }].  \label{nadact}
\end{equation}%
Introducing the absolute antisymmetric tensor $\epsilon _{\alpha \beta
\gamma \delta }$ and the effective source 3--form
\begin{equation*}
\mathcal{T}_{\ \beta }=\mathbf{T}_{\ \beta }^{\underline{\alpha }}\ \epsilon
_{\underline{\alpha }\underline{\beta }\underline{\gamma }\underline{\delta }%
}du^{\underline{\beta }}\wedge du^{\underline{\gamma }}\wedge du^{\underline{%
\delta }}
\end{equation*}%
and expressing the curvature tensor $\widehat{\mathcal{R}}_{\ \gamma }^{\tau
}=\widehat{\mathbf{R}}_{\ \gamma \alpha \beta }^{\tau }\ \mathbf{e}^{\alpha
}\wedge \ \mathbf{e}^{\beta }$ of $\ \widehat{\mathbf{\Gamma }}_{\ \beta
\gamma }^{\alpha }=\ _{\shortmid }\Gamma _{\ \beta \gamma }^{\alpha }-\ \
_{\shortmid }\widehat{\mathbf{Z}}_{\ \beta \gamma }^{\alpha }$ as $\widehat{%
\mathcal{R}}_{\ \gamma }^{\tau }=\ _{\shortmid }\mathcal{R}_{\ \gamma
}^{\tau }-\ _{\shortmid }\widehat{\mathcal{Z}}_{\ \gamma }^{\tau },$ where $%
\ _{\shortmid }\mathcal{R}_{\ \gamma }^{\tau }$ $=\ _{\shortmid }R_{\ \gamma
\alpha \beta }^{\tau }\ \mathbf{e}^{\alpha }\wedge \ \mathbf{e}^{\beta }$ is
the curvature 2--form of the Levi--Civita connection $\nabla $ and the
distorsion of curvature 2--form $\widehat{\mathcal{Z}}_{\ \gamma }^{\tau }$
is defined by $\ \widehat{\mathbf{Z}}_{\ \beta \gamma }^{\alpha },$ see (\ref%
{cdeft}) and (\ref{aux01}), we derive the equations (\ref{deinsteq})
(varying the action on components of $\mathbf{e}_{\ \beta }).$ The
gravitational field equations are represented as 3--form equations,%
\begin{equation}
\epsilon _{\alpha \beta \gamma \tau }\left( \mathbf{e}^{\alpha }\wedge
\widehat{\mathcal{R}}^{\beta \gamma }+\lambda \mathbf{e}^{\alpha }\wedge \
\mathbf{e}^{\beta }\wedge \ \mathbf{e}^{\gamma }\right) =8\pi G\mathcal{T}%
_{\ \tau },  \label{einsteq}
\end{equation}%
when%
\begin{eqnarray*}
\mathcal{T}_{\ \tau } &=&\ ^{m}\mathcal{T}_{\ \tau }+\ ^{Z}\widehat{\mathcal{%
T}}_{\ \tau }, \\
\ ^{m}\mathcal{T}_{\ \tau } &=&\ ^{m}\mathbf{T}_{\ \tau }^{\underline{\alpha
}}\epsilon _{\underline{\alpha }\underline{\beta }\underline{\gamma }%
\underline{\delta }}du^{\underline{\beta }}\wedge du^{\underline{\gamma }%
}\wedge du^{\underline{\delta }}, \\
\ ^{Z}\mathcal{T}_{\ \tau } &=&\left( 8\pi G\right) ^{-1}\widehat{\mathcal{Z}%
}_{\ \tau }^{\underline{\alpha }}\epsilon _{\underline{\alpha }\underline{%
\beta }\underline{\gamma }\underline{\delta }}du^{\underline{\beta }}\wedge
du^{\underline{\gamma }}\wedge du^{\underline{\delta }},
\end{eqnarray*}%
where $\ ^{m}\mathbf{T}_{\ \tau }^{\underline{\alpha }}=\delta \ ^{matter}%
\mathcal{S}/\delta \mathbf{e}_{\underline{\alpha }}^{\ \tau }.$ The above
mentioned equations are equivalent to the usual Einstein equations for the
Levi--Civita connection $\nabla ,$%
\begin{equation*}
\ _{\shortmid }\mathbf{R}_{\ \beta }^{\underline{\alpha }}-\frac{1}{2}(\
_{\shortmid }R+\lambda )\mathbf{e}_{\ \beta }^{\underline{\alpha }}=8\pi G\
^{m}\mathbf{T}_{\ \beta }^{\underline{\alpha }}.
\end{equation*}

The vacuum Einstein equations with cosmological constant, written in terms
of the canonical N--adapted vierbeins and normal d--connection, are%
\begin{equation}
\epsilon _{\alpha \beta \gamma \tau }\left( \mathbf{e}^{\alpha }\wedge
\widehat{\mathcal{R}}^{\beta \gamma }+\lambda \mathbf{e}^{\alpha }\wedge
\mathbf{e}^{\beta }\wedge \ \mathbf{e}^{\gamma }\right) =8\pi G\ ^{Z}%
\widehat{\mathcal{T}}_{\ \tau },  \label{veinst1}
\end{equation}%
with effective source $\ ^{Z}\widehat{\mathcal{T}}_{\ \tau }$ induced by
nonholonomic splitting by the metric tensor and its off--diagonal components
transformed into the N--connection coefficients or, in terms of the
Levi--Civita connection%
\begin{equation*}
\epsilon _{\alpha \beta \gamma \tau }\left( \mathbf{e}^{\alpha }\wedge \
_{\shortmid }\mathcal{R}^{\beta \gamma }+\lambda \mathbf{e}^{\alpha }\wedge
\mathbf{e}^{\beta }\wedge \ \mathbf{e}^{\gamma }\right) =0.
\end{equation*}%
Such formulas are necessary for encoding the vacuum field equations into
cohomological structure of quantum almost K\"{a}hler model of the Einstein
gravity, see \cite{vqg3}.

If former geometric constructions in GR and LQG were related to frame and
coordinate form invariant transforms, various purposes in geometric
modelling of physical interactions and quantization request application of
more general classes of transforms. For such generalizations, the linear
connection structure is deformed (in a unique/canonical form following well
defined geometric and physical principles) and there are considered
nonholonomic spacetime distributions. All geometric and physical information
for any data 1) $[\mathbf{g,}\ _{\shortmid }\Gamma _{\ \alpha \beta
}^{\gamma }(\mathbf{g})]$ are transformed equivalently for canonical
constructions with 2) $[\mathbf{g}=\ ^{L}\mathbf{g,}\ \ ^{L}\mathbf{N},$ $\
\widehat{\mathbf{\Gamma }}_{\ \alpha \beta }^{\gamma }(\ ^{L}\mathbf{g})],$
which allows us to provide an effective Lagrange--Finsler like
interpretation of the Einstein gravity, or 3) $[\ ^{L}\mathbf{\theta },\
_{\theta }\widehat{\mathbf{\Gamma }}_{\ \alpha \beta }^{\gamma }=\widehat{%
\mathbf{\Gamma }}_{\ \alpha \beta }^{\gamma },\mathbf{J(\ ^{L}\mathbf{g})}],$
for an almost K\"{a}hler model of general relativity. The Einstein equations
for the Levi--Civita connection can be written in a more ''simple'' form
following the approach 1). They are redefined equivalently using
corresponding distorsions tensors and data 2) and 3) but in a more
cumbersome form (a similar ''sophistication'' holds if the Barbero variables %
\cite{barb1,barb2} are introduced instead of the Ashtekar ones in order to
get real constraints in the effective phase space for LQG).

Following a program oriented to a Fedosov like quantization of general
relativity, it is important to work with models of type 3) and related
models of type 2). In such cases, we positively work with almost K\"{a}hler
spaces for effective Lagrange--Finsler geometry which are more simple than
those derived for almost Hermitian geometries; see, for instance, some
constructions and discussions related to generalized Lagrange geometries in %
\cite{vlqgdq2,vrfg}.

\section{N--adapted (3+1)--Splitting in GR}

In this section, the variables that are used in LQG are re--defined in a
form adapted to a nonlinear connection (N--connection) structure. We
consider that any spacetime $\mathbf{V}$ enabled with a (pseudo) Riemannian
metric $\mathbf{g}$ (\ref{gpsm}) is provided with two foliations: the first
one reflects the semispray structure $S$ (\ref{sspray}) defined by an
effective regular Lagrangian $L,$ with associated canonical N--connection $\
^{L}\mathbf{N}$ (\ref{clnc}), inducing a nonholonomic 2+2 splitting; the
second one is described in terms of space--like tree dimensional surfaces $\
^{3}\mathbf{\Sigma }$ (for simplicity, we assume that such surfaces are with
no boundaries), which defines a 3+1 splitting.

\subsection{The Palatini action in N--adapted ADM variables}

A metric $\ \mathbf{g}$ (\ref{gpsm}) is equivalently transformed into a
d--metric $\ ^{L}\mathbf{g}$ (\ref{lfsm}) if we perform a frame (vielbein)
transform
\begin{equation}
\mathbf{e}_{\alpha }=\mathbf{e}_{\alpha }^{\ \underline{\alpha }}\partial _{%
\underline{\alpha }}\mbox{ and }\mathbf{e}_{\ }^{\beta }=\mathbf{e}_{\
\underline{\beta }}^{\beta }du^{\underline{\beta }},  \label{ftnc}
\end{equation}%
with coefficients
\begin{eqnarray}
\mathbf{e}_{\alpha }^{\ \underline{\alpha }}(u) &=&\left[
\begin{array}{cc}
e_{i}^{\ \underline{i}}(u) & \ ^{L}N_{i}^{b}(u)e_{b}^{\ \underline{a}}(u) \\
0 & e_{a}^{\ \underline{a}}(u)%
\end{array}%
\right] ,  \label{vt1} \\
\mathbf{e}_{\ \underline{\beta }}^{\beta }(u) &=&\left[
\begin{array}{cc}
e_{\ \underline{i}}^{i\ }(u) & -\ ^{L}N_{k}^{b}(u)e_{\ \underline{i}}^{k\
}(u) \\
0 & e_{\ \underline{a}}^{a\ }(u)%
\end{array}%
\right] ,  \label{vt2}
\end{eqnarray}%
being linear on $N_{i}^{a}.$ So, with respect to a local coordinate basis $%
du^{\alpha }=(dx^{i},dy^{a}),$ any metric can be parametrized in the form
\begin{equation}
\ ^{L}\mathbf{g}=\ ^{L}\underline{\mathbf{g}}_{\alpha \beta }\left( u\right)
du^{\alpha }\otimes du^{\beta },  \label{metr}
\end{equation}%
where%
\begin{equation}
\ ^{L}\underline{\mathbf{g}}_{\alpha \beta }=\left[
\begin{array}{cc}
\ ^{L}g_{ij}+\ ^{L}N_{i}^{a}\ ^{L}N_{j}^{b}\ ^{L}g_{ab} & \ ^{L}N_{j}^{e}\
^{L}g_{ae} \\
\ ^{L}N_{i}^{e}\ ^{L}g_{be} & \ ^{L}g_{ab}%
\end{array}%
\right] ,  \label{ansatz}
\end{equation}%
for%
\begin{equation}
\ ^{L}\underline{\mathbf{g}}_{\alpha \beta }=\mathbf{e}_{\alpha }^{\ \alpha
^{\prime }}\mathbf{e}_{\beta }^{\ \beta ^{\prime }}\eta _{\alpha ^{\prime
}\beta ^{\prime }}=\mathbf{e}_{\alpha }^{\ \underline{\alpha }}\mathbf{e}%
_{\beta }^{\ \underline{\beta }}\mathbf{g}_{\underline{\alpha }\underline{%
\beta }},  \label{orthmdec}
\end{equation}%
with $\eta _{\alpha ^{\prime }\beta ^{\prime }}=diag[-1,1,1,1].$ These
formulas are adapted to a nonholonomic 2+2 splitting.

The tetrad variables $\mathbf{e}_{\ \beta ^{\prime }}^{\beta }$ from (\ref%
{orthmdec}) and a $so(1,3)$--valued connection $\Gamma _{\ \beta ^{\prime
}\alpha }^{\alpha ^{\prime }}$ (not necessarily torsion free, but metric
compatible) subjected to the vierbein transform rule%
\begin{equation}
\ \Gamma _{\ \gamma ^{\prime }\beta ^{\prime }}^{\alpha ^{\prime }}=\mathbf{e%
}_{\alpha }^{\ \alpha ^{\prime }}\mathbf{e}_{\ \gamma ^{\prime }}^{\gamma }%
\mathbf{e}_{\ \beta ^{\prime }}^{\beta }\Gamma _{\ \gamma \beta }^{\alpha }+%
\mathbf{e}_{\gamma }^{\ \alpha ^{\prime }}\mathbf{e}_{\ \beta ^{\prime
}}^{\beta }\mathbf{e}_{\beta }(\mathbf{e}_{\ \gamma ^{\prime }}^{\gamma }),
\label{ftrlc}
\end{equation}%
where $\alpha ^{\prime },\beta ^{\prime },...$ are considered as internal
indices, can be used for a generalized Palatini approach to gravity \cite%
{asht} with the action%
\begin{equation}
^{P}\mathcal{S}[\mathbf{e}_{\ \beta ^{\prime }}^{\beta },\Gamma _{\ \beta
^{\prime }\alpha }^{\alpha ^{\prime }}]=\frac{1}{2\kappa }\int\nolimits_{%
\mathbf{V}}\delta ^{4}u\ \sqrt{|\ \mathbf{g}|}\mathbf{e}_{\ \alpha ^{\prime
}}^{\alpha }\mathbf{e}_{\ \beta ^{\prime }}^{\beta }\left( \mathcal{R}%
_{\quad \alpha \beta }^{\alpha ^{\prime }\beta ^{\prime }}+\frac{1}{2\beta }%
\epsilon _{\quad \tau ^{\prime }\nu ^{\prime }}^{\alpha ^{\prime }\beta
^{\prime }}\mathcal{R}_{\quad \alpha \beta }^{\tau ^{\prime }\nu ^{\prime
}}\right) .  \label{paction}
\end{equation}%
In the above formulas, $\epsilon _{\quad \tau ^{\prime }\nu ^{\prime
}}^{\alpha ^{\prime }\beta ^{\prime }}$ is the internal Levi--Civita symbol,
$\kappa =8\pi G/c^{3}=8\pi l_{p}^{2}/\hbar $ (here $G$ is the Newton
gravitational constant, $c$ is the light velocity constant, $l_{p}\sim
10^{-33}cm$ is the Planck scale and $\hbar =h/2\pi $ is the Planck constant)
$\beta $ is the Barbero--Immirzi parameter \cite{barb1,barb2,immirzi} and
the $so(1,3)$--valued curvature 2--form is computed%
\begin{equation}
\mathcal{R}_{\quad \beta ^{\prime }\alpha \beta }^{\alpha ^{\prime
}}\doteqdot \mathcal{D}_{\alpha }\Gamma _{\ \beta ^{\prime }\beta }^{\alpha
^{\prime }}-\mathcal{D}_{\beta }\Gamma _{\ \beta ^{\prime }\alpha }^{\alpha
^{\prime }}=\mathbf{e}_{\alpha }\Gamma _{\ \beta ^{\prime }\beta }^{\alpha
^{\prime }}-\mathbf{e}_{\beta }\Gamma _{\ \beta ^{\prime }\alpha }^{\alpha
^{\prime }}+\Gamma _{\ \tau ^{\prime }\alpha }^{\alpha ^{\prime }}\wedge
\Gamma _{\ \beta ^{\prime }\beta }^{\tau ^{\prime }}  \label{curv}
\end{equation}%
for the covariant derivative $\mathcal{D}_{\alpha },$ defined by $\Gamma _{\
\beta ^{\prime }\alpha }^{\alpha ^{\prime }},$ acting both on spacetime and
internal indices.

The vacuum gravitational field equations are obtained by varying the action (%
\ref{paction}) with respect to $\mathbf{e}_{\ \beta ^{\prime }}^{\beta }$
and $\Gamma _{\ \beta ^{\prime }\alpha }^{\alpha ^{\prime }}.$ If we take $%
\Gamma _{\ \beta ^{\prime }\alpha }^{\alpha ^{\prime }}=\ _{\shortmid
}\Gamma _{\ \beta ^{\prime }\alpha }^{\alpha ^{\prime }},$ we get the vacuum
Einstein equations in GR. For $\Gamma _{\ \beta ^{\prime }\alpha }^{\alpha
^{\prime }}=\ \ \widehat{\mathbf{\Gamma }}_{\ \beta ^{\prime }\alpha
}^{\alpha ^{\prime }},$ see (\ref{cdeft}), we have $^{P}\mathcal{S}[\mathbf{e%
}_{\ \beta ^{\prime }}^{\beta },\ \widehat{\mathbf{\Gamma }}_{\ \beta
^{\prime }\alpha }^{\alpha ^{\prime }}]=\ ^{gr}\mathcal{S}[\mathbf{e,}%
\widehat{\mathbf{\Gamma }}],$ see (\ref{nadact}), which results in the
fields equations (\ref{veinst1}) and Cartan's first system of equations
\begin{equation*}
\widehat{\mathcal{D}}_{[\alpha }\mathbf{e}_{\beta ]}^{\ \alpha ^{\prime }}=%
\widehat{\mathbf{T}}_{\alpha \beta }^{\alpha ^{\prime }},
\end{equation*}%
with effective d--torsion $\widehat{\mathbf{T}}_{\alpha \beta }^{\alpha
^{\prime }}$ (\ref{cdtors}) defined canonically by the metric coefficients $%
\ \mathbf{g}=\ ^{L}\underline{\mathbf{g}}_{\alpha \beta }.$ The linearity,
N--connection splitting (not N--adapted, for the Levi--Civita case) and
metric compatibility can be provided if
\begin{equation}
\bigtriangledown _{\alpha }\mathbf{e}_{\alpha }^{\ \alpha ^{\prime
}}=0\Longrightarrow \ _{\shortmid }\Gamma _{\ \beta ^{\prime }\alpha
}^{\alpha ^{\prime }}=\mathbf{e}_{\ \beta ^{\prime }}^{\beta }\left( \mathbf{%
e}_{\tau }^{\ \alpha ^{\prime }}\ _{\shortmid }\Gamma _{\ \alpha \beta
}^{\tau }-\mathbf{e}_{\alpha }(\mathbf{e}_{\beta }^{\ \alpha ^{\prime
}})\right)  \label{auxc1}
\end{equation}%
and
\begin{equation*}
\widehat{\mathbf{D}}_{\alpha }\mathbf{e}_{\alpha }^{\ \alpha ^{\prime
}}=0\Longrightarrow \ \ \ \widehat{\mathbf{\Gamma }}_{\ \beta ^{\prime
}\alpha }^{\alpha ^{\prime }}=\mathbf{e}_{\ \beta ^{\prime }}^{\beta }\left(
\mathbf{e}_{\tau }^{\ \alpha ^{\prime }}\ \ \ \widehat{\mathbf{\Gamma }}_{\
\alpha \beta }^{\tau }-\mathbf{e}_{\alpha }(\mathbf{e}_{\beta }^{\ \alpha
^{\prime }})\right) .
\end{equation*}%
It should be noted that with respect to N--adapted bases (\ref{dder}) and (%
\ref{ddif}) both linear connections $\bigtriangledown $ and $\widehat{%
\mathbf{D}}$ are given by the same coefficients (\ref{cdcc}). This may
simplify various coordinate and N--adapted local computations. Nevertheless,
it should be noted that even both type of metric compatible connections are
defined canonically by the same metric $\ \mathbf{g}=\ ^{L}\mathbf{g}%
_{\alpha \beta },$ they are really different with different properties and
laws of frame and coordinate transforms (for the normal d--connection, there
is an induced nontrivial torsion structure (\ref{cdtors}) defined by the
nonholonomy relations (\ref{anhrel})).\footnote{%
A formal equality of the coefficients of different connections can be
obtained with respect to some local bases because the linear connections are
not tensor geometric objects.}

\subsection{N--adapted 3+1 splitting and the Hamilton formalism}

To carry out a 3+1 fibrated Hamilton analysis of action (\ref{paction}), we
suppose for simplicity that the spacetime $\mathbf{V}$ is topologically $\
^{3}\mathbf{\Sigma \times }\mathbb{R}.$ The fibration is parametrized by a
smooth function $t$ and a time--evolution vector field $t^{\alpha }$ such
that $t^{\alpha }(dt)_{\alpha }=1$ (it is a d--vector field if we work with
N--adapted bases; in such cases we shall use a boldface symbol, i.e. $%
\mathbf{t}^{\alpha }).$ Using the unit normal d--vector $\mathbf{n}^{\alpha
} $ of $^{3}\mathbf{\Sigma ,}$ we decompose%
\begin{equation}
\mathbf{t}^{\alpha }=\ _{l}\mathbf{Nn}^{\alpha }+\ _{s}\mathbf{N}^{\alpha },
\label{laps}
\end{equation}%
where $\ _{l}\mathbf{N}$ is the lapse function and $_{s}\mathbf{N}^{\alpha }$
is the shift d--vector.\footnote{%
we put left low labels to such symbols in order to not confuse them with the
symbol $\mathbf{N}$ for a N--connection and to preserve the style of
notations from the ADM formalism but, in our case, adapted also to a 2+2
splitting} It is convenient to work with a partial gauge fixing, when $\eta
_{\alpha ^{\prime }\beta ^{\prime }}\mathbf{n}^{\alpha ^{\prime }}\mathbf{n}%
^{\beta ^{\prime }}=-1$ for $\mathbf{n}_{\alpha ^{\prime }}\equiv \mathbf{n}%
_{\alpha }\mathbf{e}_{\ \alpha ^{\prime }}^{\alpha }.$ The primed indices $%
\alpha ^{\prime },\beta ^{\prime },...$ are considered as internal ones
which split $\alpha ^{\prime }=(0^{\prime },I^{\prime }),$ $\beta ^{\prime
}=(0^{\prime },J^{\prime }),...$ for $I^{\prime },J^{\prime },...=1,2,3$
used for the internal space on $^{3}\mathbf{\Sigma .}$

We consider a triad $e_{I}$ (a set of three 1--forms defining a 3
dimensional coframe in each point in $\ ^{3}\mathbf{\Sigma ),}$ when $%
e_{\alpha }=(e_{0},e_{I})$ is defined on $\mathbf{V.}$ A three dimensional
metric $\mathbf{q}_{IJ}$ on $\ ^{3}\mathbf{\Sigma }$ is parametrized
\begin{equation*}
\mathbf{q}_{IJ}=\mathbf{e}_{I}^{\ I^{\prime }}\mathbf{e}_{J}^{\ J^{\prime
}}\delta _{I^{\prime }J^{\prime }},
\end{equation*}%
where $\delta _{I^{\prime }J^{\prime }}$ is the Kronecker symbol and indices
of type $I,J,...$ are used to denote the abstract indices of space $\ ^{3}%
\mathbf{\Sigma .}$ The orthonormal co--triad is defined by
\begin{equation*}
\mathbf{e}_{I}^{\ I^{\prime }}\doteqdot \mathbf{e}_{\alpha }^{\ \alpha
^{\prime }}q_{\alpha ^{\prime }}^{I^{\prime }}q_{I}^{\alpha },
\end{equation*}%
for the internal and spacetime projection maps denoted respectively $%
q_{I}^{\alpha }$ and $q_{\alpha ^{\prime }}^{I^{\prime }},$ which reduces
the internal gauge group $SO(1,3)$ to $SO(3)$ living invariant $\mathbf{n}%
^{\alpha ^{\prime }}.$ Such formulas define a 3+1 spacetime fibration which
can be adapted to a N--connection structure $\ ^{L}\mathbf{N}$ (\ref{clnc})
induced by an effective Lagrangian $L$ if we consider instead of arbitrary $%
e_{\alpha }$ and $e^{\beta }$ just only N--adapted vielbeins $\mathbf{e}%
_{\alpha }=(\mathbf{e}_{0},\mathbf{e}_{I})$ and $\mathbf{e}_{\ }^{\beta }=(%
\mathbf{e}_{\ }^{0},\mathbf{e}_{\ }^{I})$ (respectively related by
non--degenerated and signature preserving frame transforms with (\ref{dder})
and (\ref{ddif})). For a ''double'' 2+2 and 3+1 spacetime splitting, adapted
to an effective $L,$ we shall use 'boldface' symbols and left labels of type
$\ ^{L}\mathbf{q}_{IJ}.$\footnote{%
Some readers may consider the system of notations to be quite sophisticate.
Nevertheless, such a system is necessary if we work with ''double''
fibrations and abstract/ coordinate index constructions adapted to different
effective Lagrange--Finsler, almost K\"{a}hler and Riemannian structures.
The final geometrical results do not depend on the type of splitting and
formalism applied for proofs but we have to use in parallel three "geometric
languages'' (from loop, almost symplectic and nonholonomic manifolds
geometries) in order to present and "translate" the results to some
separated communities of researches working with different methods of
quantization of gravity and field theories.}

Let us consider a metric compatible affine connection (in general, it may be
not a d--connection) $D=\{\Gamma _{\ \beta \gamma }^{\alpha }\},$ when $D%
\mathbf{g}=0$ on $\mathbf{V}.$ We have a $so(1,3)$--valued connection 1-form
\begin{equation*}
\Gamma _{\ \gamma ^{\prime }}^{\alpha ^{\prime }}=\Gamma _{\ \gamma ^{\prime
}\beta ^{\prime }}^{\alpha ^{\prime }}\mathbf{e}_{\ }^{\beta ^{\prime }}
\end{equation*}%
with coefficients defined by formulas (\ref{ftrlc}). Using the projection
maps $q_{I}^{\alpha }$ and $q_{\alpha ^{\prime }}^{I^{\prime }}$ and the
anti--symmetric Levi--Civita symbol $\epsilon ^{\alpha ^{\prime }\beta
^{\prime }\gamma ^{\prime }\tau ^{\prime }},$ we construct two $so(3)$%
--valued 1--forms induced by \ $\Gamma _{\ \beta \gamma }^{\alpha }$ on $\
^{3}\mathbf{\Sigma :}$%
\begin{equation}
\Gamma _{\ I}^{I^{\prime }}\doteqdot \frac{1}{2}q_{I}^{\alpha }q_{\alpha
^{\prime }}^{I^{\prime }}\epsilon _{\quad \gamma ^{\prime }\tau ^{\prime
}}^{\alpha ^{\prime }\beta ^{\prime }}\mathbf{n}_{\beta ^{\prime }}\Gamma
_{\quad \alpha }^{\gamma ^{\prime }\tau ^{\prime }}\mbox{ and }K_{\
I}^{I^{\prime }}\doteqdot q_{\alpha ^{\prime }}^{I^{\prime }}q_{I}^{\alpha }%
\mathbf{n}_{\beta ^{\prime }}\Gamma _{\quad \alpha }^{\alpha ^{\prime }\beta
^{\prime }},  \label{so3cf}
\end{equation}%
called respectively the spin connection and the extrinsic curvature on shell.

The modern LQG \cite{rov,thiem1,asht,smol} is formulated for a class of
objects $(\ _{\shortmid }\Gamma _{\ I}^{I^{\prime }},\ _{\shortmid }K_{\
I}^{I^{\prime }})$ induced following formulas (\ref{ftrlc}) and (\ref{so3cf}%
) by the Levi--Civita connection, i.e. when $\Gamma _{\ \alpha \beta
}^{\gamma }=\ _{\shortmid }\Gamma _{\ \alpha \beta }^{\gamma }(\mathbf{g}).$
One considers a classical phase space and Hamiltonian formalism for
variables $(\ _{\shortmid }A_{\ I}^{I^{\prime }},\widetilde{E}_{I^{\prime
}}^{\quad I})$ when the configuration and conjugate momentum are defined
respectively%
\begin{eqnarray}
\ _{\shortmid }A_{\ I}^{I^{\prime }} &\doteqdot &\ _{\shortmid }\Gamma _{\
I}^{I^{\prime }}+\beta \ _{\shortmid }K_{\ I}^{I^{\prime }},  \label{ashtcon}
\\
\widetilde{E}_{I^{\prime }}^{\quad I} &=&\beta \widetilde{P}_{I^{\prime
}}^{\quad I}\doteqdot (2\kappa \beta )^{-1}\widetilde{\epsilon }%
^{IJK}\epsilon _{I^{\prime }J^{\prime }K^{\prime }}\mathbf{e}_{J}^{\
J^{\prime }}\mathbf{e}_{K}^{\ K^{\prime }}=(\kappa \beta )^{-1}\sqrt{q}%
\mathbf{e}_{\ I^{\prime }}^{I\ },  \notag
\end{eqnarray}%
where $q\doteqdot \det |\mathbf{q}_{IJ}|=(\kappa \beta )^{3}\det
|P_{I^{\prime }}^{\quad I}|$ and ''tilde'' is used for density objects. In
terms of such variables, we can define $\ _{\shortmid }\check{\Gamma}_{\
I}=\ _{\shortmid }\Gamma _{\ I}^{I^{\prime }}\check{\tau}_{I^{\prime }},$
where $\left( \check{\tau}_{I^{\prime }}\right) _{J^{\prime }K^{\prime
}}=\epsilon _{J^{\prime }I^{\prime }K^{\prime }}$ are the generators of $%
so(3)$ (or equivalently, of $su(2)$ in the adjoint representation) if the
structure constants are chosen to be $\epsilon _{J^{\prime }I^{\prime
}K^{\prime }}$ and write the conditions (\ref{auxc1}) as%
\begin{equation*}
G_{I^{\prime }}=\ _{\shortmid }D_{I}E_{I^{\prime }}^{\quad I}\doteqdot
\mathbf{e}_{I}E_{I^{\prime }}^{\quad I}+\epsilon _{I^{\prime }J^{\prime
}}^{\quad K^{\prime }}\ _{\shortmid }\Gamma _{\ I}^{J^{\prime }}\
E_{K^{\prime }}^{\quad I}=0,
\end{equation*}%
or, equivalently, as
\begin{equation*}
\ _{\shortmid }^{\beta }D_{I}\ ^{\beta }E_{I^{\prime }}^{\quad I}\doteqdot 0,
\end{equation*}%
for $(\ _{\shortmid }K_{\ I}^{I^{\prime }},E_{I^{\prime }}^{\quad
I})\rightarrow (\ _{\shortmid }^{\beta }K_{\ I}^{I^{\prime }}\doteqdot \beta
\ _{\shortmid }K_{\ I}^{I^{\prime }},\ ^{\beta }E_{I^{\prime }}^{\quad
I}\doteqdot E_{I^{\prime }}^{\quad I}/\beta ).$ These equations suggested to
consider $\ _{\shortmid }A_{\ I}^{I^{\prime }}$ from (\ref{ashtcon}) as a
new linear connection called the Ashtekar connection, or the
Ashtekar--Barbero connection.\footnote{%
Following historical reasons, we might call this connection the Sen --
Ashtekar -- Immirzi -- Barbero connection (the Sen connection \cite{sen}
arises for $\beta =\pm i,G_{I}=0,$ when the last condition is called the
Gauss constraint: the Ashtekar connection \cite{asht1,asht2} is for $\beta
=\pm i,$ the Immirzi connection \cite{immirzi} is for complex $\beta $ and
the Barbero connection \cite{barb1,barb2} is for real $\beta ).$} In such
new Ashtekar variables, the action (\ref{paction}), see details in \cite%
{rov,thiem1}, results in the Hamiltonian density
\begin{equation*}
\mathit{H}_{tot}=\Lambda ^{I^{\prime }}G_{I^{\prime }}+\ _{s}N^{I}C_{I}+\
_{l}NC,
\end{equation*}%
where $\Lambda ^{I^{\prime }}\doteqdot -\frac{1}{2}\epsilon _{\ }^{I^{\prime
}J^{\prime }K^{\prime }}\ _{\shortmid }\Gamma _{J^{\prime }K^{\prime }0},\
_{s}N^{I}$ and $\ _{l}N$ are Lagrange multiples (contrary to (\ref{laps}),
we do not use boldface symbols because $\ _{\shortmid }\Gamma _{\
I}^{J^{\prime }}$ is not a d--connection) and the constraints are
\begin{eqnarray}
G_{I^{\prime }} &=&\ _{\shortmid }D_{I}\widetilde{P}_{I^{\prime }}^{\quad
I}\doteqdot \mathbf{e}_{I}\ \widetilde{P}_{I^{\prime }}^{\quad I}+\epsilon
_{I^{\prime }J^{\prime }}^{\quad K^{\prime }}\ _{\shortmid }A_{\
I}^{J^{\prime }}\ \widetilde{P}_{K^{\prime }}^{\quad I},  \label{constr} \\
C_{I} &=&\widetilde{P}_{I^{\prime }}^{\quad J}\ _{\shortmid
}F_{IJ}^{I^{\prime }}-\beta ^{-1}(1+\beta ^{2})\ _{\shortmid }K_{\
I}^{I^{\prime }}G_{I^{\prime }},  \notag \\
C &=&(\kappa \beta ^{2}/2\sqrt{q})\widetilde{P}_{I^{\prime }}^{\quad I}%
\widetilde{P}_{J^{\prime }}^{\quad J}\left[ \epsilon _{\ \quad K^{\prime
}}^{I^{\prime }J^{\prime }}\ _{\shortmid }F_{IJ}^{K^{\prime }}-2(1+\beta
^{2})\ _{\shortmid }K_{\ [I}^{I^{\prime }}\ _{\shortmid }K_{\ J]}^{J^{\prime
}}\right]  \notag \\
&&+\kappa (1+\beta ^{2})\mathbf{e}_{I}(\sqrt{q}\widetilde{P}_{I^{\prime
}}^{\quad I})G^{I^{\prime }},  \notag
\end{eqnarray}%
where the $so(3)$--valued curvature 2--form of $\ _{\shortmid }A_{\
I}^{I^{\prime }}$ is defined as in Yang--Mills theory,%
\begin{equation*}
\ _{\shortmid }F_{IJ}^{K^{\prime }}\doteqdot \mathbf{e}_{I}(\ _{\shortmid
}A_{\ J}^{K^{\prime }})-\mathbf{e}_{J}(\ _{\shortmid }A_{\ I}^{K^{\prime
}})+\epsilon _{\ I^{\prime }J^{\prime }}^{K^{\prime }}\ _{\shortmid }A_{\
I}^{I^{\prime }}\ _{\shortmid }A_{\ J}^{J^{\prime }}.
\end{equation*}

Introducing the Hamiltonian
\begin{equation*}
H\doteqdot \int\nolimits_{\mathbf{\Sigma }}d^{3}x_{\Sigma }\ \mathit{H}%
_{tot},
\end{equation*}%
where $d^{3}x_{\Sigma }$ is the volume element on the hypersurface $\ ^{3}%
\mathbf{\Sigma ,}$ we get a symplectic structure on the classical phase
space with the Poisson brackets of type
\begin{equation}
\{\ _{\shortmid }A_{\ J}^{J^{\prime }}(\ ^{1}x),\widetilde{P}_{I^{\prime
}}^{\quad I}(\ ^{2}x)\}=\delta _{I^{\prime }}^{J^{\prime }}\delta
_{J}^{I}\delta (\ ^{1}x-\ ^{2}x),  \label{poisson1}
\end{equation}%
for any two points $\ ^{1}x$ and $\ ^{2}x$ on $\ ^{3}\mathbf{\Sigma .}$ The
constraints algebra for (\ref{constr}) is closed under such a Poisson
structure, i.e. all constraints are of first class following the Dirac
approach to quantization of constrained systems and the hamiltonian $H$ is
the linear combination of the constraints functions.

The evolution equations
\begin{equation}
\mathcal{L}_{t}\ _{\shortmid }A_{\ I}^{I^{\prime }}=\{\ _{\shortmid }A_{\
I}^{I^{\prime }},H\}\mbox{ and }\mathcal{L}_{t}\ \widetilde{P}_{K^{\prime
}}^{\quad I}=\{\ \widetilde{P}_{K^{\prime }}^{\quad I},H\},  \label{eveq}
\end{equation}%
where $\mathcal{L}_{t}$ is the Lie derivative with respect to the time--like
direction, together with the constraints equations (\ref{constr}), are
completely equivalent to the vacuum Einstein equations.\footnote{%
for simplicity, we do not consider here matter field sources, which can be
also introduced into ADM and LQG formalisms, see \cite{rov,thiem1}} In
result of such constructions, GR is cast as a dynamical theory of
connections with a compact structure group. Together with the 'Master
Constraint Project' \cite{thiem,thiem2}, this allows to solve the
constraints problem and perform a loop quantization of gravity.

\section{Fedosov--Ashtekar N--adapted Operators}

We provide a nonholonomic generalization of the Ashtekar--Barbero variables %
\cite{asht1,asht2,barb1,barb2} and define the corresponding star product and
Fedosov--Ashtekar d--connection. The Fedosov's deformation quantization \cite%
{fed1,fed2} generalized for almost K\"{a}hler geometries \cite%
{karabeg1,sternh1} will be applied to (pseudo) Riemannian manifolds
parametized in Lagrange--Finsler variables.

\subsection{N--adapted Ashtekar--Barbero variables}

In section \ref{ssakgr}, \ we formulated an almost K\"{a}hler model of GR in
variables $[\ ^{L}\mathbf{\theta },\ _{\theta }\widehat{\mathbf{\Gamma }}_{\
\alpha \beta }^{\gamma }=\widehat{\mathbf{\Gamma }}_{\ \alpha \beta
}^{\gamma },\mathbf{J(\ ^{L}\mathbf{g})}],$ where the almost symplectic form
$\ ^{L}\mathbf{\theta }$ is given by formula (\ref{asymstr}), the normal
connection $_{\theta }\widehat{\mathbf{\Gamma }}_{\ \alpha \beta }^{\gamma }=%
\widehat{\mathbf{\Gamma }}_{\ \alpha \beta }^{\gamma }$ has N--adapted
coefficients (\ref{cdcc}) and the almost complex structure $\mathbf{J(\ ^{L}%
\mathbf{g})}$ is defined by the tensor (\ref{acstr}) with a corresponding
Neijenhuis tensor $\ ^{\mathbf{J}}\mathbf{\Omega }$ (\ref{neijt}), all
constructions being adapted to the N--connection structure $\ ^{L}\mathbf{N}$
(\ref{clnc}).

Let us consider a general frame $e_{\nu }$ and corresponding (dual) coframe $%
e^{\mu }$ on a spacetime $V$ enabled with metric $g,$ almost complex, $%
J(e_{\beta })=J_{\ \beta }^{\alpha }e_{\alpha },$ and symplectic, $\theta ,$
structures in compatible forms in the following sense: $\theta (JX,JY)=$ $%
\theta (X,Y)$ and $g(X,Y)=\theta (JX,Y)$ for any vector fields $X,Y\in TV.$
For $\theta _{\alpha \beta } \doteqdot \theta (e_{\alpha },e_{\beta })$ and $%
g_{\alpha \beta }\doteqdot g(e_{\alpha },e_{\beta }),$ and introducing the
corresponding inverse matrices $g^{\alpha \beta }$ and $\theta ^{\alpha
\beta },$ we have $J_{\ \beta }^{\alpha }=g_{\beta \gamma }\theta ^{\gamma
\alpha }=g^{\alpha \gamma }\theta _{\gamma \beta },$ which allows us to
compute the coefficients of the Neijenhuis tensor $^J\Omega =\ ^J \Omega
_{\alpha \beta }^{\gamma }\ e_{\gamma }$ for this complex structure,%
\footnote{%
we can use the formula (\ref{neijt}) but for not ''boldfaced'' geometric
objects}
\begin{equation}
\ ^J\Omega _{\alpha \beta }^{\gamma }=(e_{\tau }J_{\ \alpha }^{\gamma })J_{\
\beta }^{\tau }-(e_{\tau }J_{\ \beta }^{\gamma })J_{\ \alpha }^{\tau
}+(e_{\alpha }J_{\ \beta }^{\tau }-e_{\beta }J_{\ \alpha }^{\tau })J_{\ \tau
}^{\gamma }.  \label{ntcoef}
\end{equation}%
An affine connection $D=\{\Gamma _{\ \alpha \beta }^{\gamma }\}$ respects
(i.e. it is compatible) the almost--complex structure $J,$ if $D_{\tau }J_{\
\alpha }^{\gamma }=0.$ One holds true a result due to \cite{yano} stating
that the torsion $T_{\beta \gamma }^{\alpha }$ of a compatible with $J$
linear connection $D$ must satisfy the condition%
\begin{equation}
4T_{\beta \gamma }^{\alpha }=-\mathbf{\ }^{\mathbf{J}}\Omega _{\alpha \beta
}^{\gamma }.  \label{compcond}
\end{equation}%
This formula was applied for generalizing the Fedosov's results to almost K%
\"{a}hler deformation quantization in Ref. \cite{karabeg1} and discussed in
Refs. \cite{vqg1,vqg2,vqg3,vqg4} for applications in quantum
Lagrange--Finsler geometry and deformation quantization of gravity. Here, we
note that working with Lagrange--Finsler variables on (pseudo) Riemannian
spaces, or in a more general case, on metric--affine and Riemann--Cartan
spaces admitting such Lagrange--Finsler modelling, we can restrict our
considerations to canonical almost K\"{a}hler models when all geometric
objects are N--adapted and compatible with the almost complex structure. The
restriction (\ref{compcond}) becomes not crucial for the related
Lagrange--Finsler d--connections and d--tensors, which allows us to perform
a deformation quantization in N--adapted form for the corresponding
canonical objects.

In order to establish a ''bridge'' between LQG and DQ of the Einstein
gravity, we have to consider possible relations between the connection and
frame variables of both theories. In a canonical way, having prescribed an
additional almost complex structure $J$ on a spacetime $V,$ we may try to
generalize the Ashtekar variables $\ _{\shortmid }A_{\ I}^{I^{\prime }}$ (%
\ref{ashtcon}) to the coefficients of an affine connection $D=\{\Gamma _{\
\alpha \beta }^{\gamma }\}$ satisfying the conditions (\ref{compcond}). In
such a case, we are able to quantize a generalized gravity model but the
constructions depend on an additional tensor object $J$ and related torsion.
Nevertheless, as we emphasized in the previous sections, using
Lagrange--Finsler variables, we are able to derive canonical almost K\"{a}%
hler geometric objects in N--adapted form, when all structures, including
the almost complex tensor $\mathbf{J(\ ^{L}\mathbf{g})}$ (\ref{acstr}) and
the corresponding Neijenhuis tensor $\ ^{\mathbf{J}}\mathbf{\Omega (\ ^{L}%
\mathbf{g})}$ (\ref{neijt}) are completely defined by the metric structure $%
\mathbf{g}=\ ^{L}\mathbf{g}$ (\ref{lfsm}). In this case, all classical data
from GR can be equivalently encoded into a nonholonomic almost K\"{a}hler
model and inversely.

Our idea is to generalize the Ashtekar connection $\ _{\shortmid }A_{\
I}^{I^{\prime }}$ from LQG to a d--connection $\ ^{A}\mathcal{D}=\{\mathbf{A}%
_{\ I}^{I^{\prime }}\},$
\begin{equation}
\ _{\shortmid }A_{\ I}^{I^{\prime }}\rightarrow \ \mathbf{A}_{\
I}^{I^{\prime }}=\ _{\shortmid }A_{\ I}^{I^{\prime }}+\ _{\shortmid }^{A}%
\mathbf{Z}_{\ I}^{I^{\prime }},  \label{ndashc}
\end{equation}%
(we call $\mathbf{A}_{\ I}^{I^{\prime }}$ the Ashtekar d--connection, or,
equivalently, the nonholonomic deformation of the Ashtekar--Barbero
connection $\ _{\shortmid }A_{\ I}^{I^{\prime }})$ where the deformation
d--tensor $\ _{\shortmid }\mathbf{Z}_{\ I}^{I^{\prime }}$ defines the
torsion of $\ \mathbf{A}_{\ I}^{I^{\prime }}$ to be canonically induced by $%
\mathbf{\ ^{L}\mathbf{g}}$ and satisfy the condition (\ref{compcond}), i.e.
\begin{equation}
4\ ^{A}\mathbf{T}_{\alpha \beta }^{\alpha ^{\prime }}=-\mathbf{\ }^{\mathbf{J%
}}\mathbf{\Omega }_{\alpha \beta }^{\alpha ^{\prime }}\mbox{\ for \ }\ ^{A}%
\mathcal{D}_{[\alpha }\mathbf{e}_{\beta ]}^{\ \alpha ^{\prime }}=\ ^{A}%
\mathbf{T}_{\alpha \beta }^{\alpha ^{\prime }}.  \label{torneij}
\end{equation}%
This will allow us to apply the formalism of Fedosov quantization in LQG
with nonholonomically deformed Ashtekar variables. In Appendix, we sketch
the method of computation and provide explicit formulas for the deformation
tensor $\ _{\shortmid }^{A}\mathbf{Z}_{\ I}^{I^{\prime }}$ (\ref{ashtdt})
which is also uniquely defined by $\mathbf{\ ^{L}\mathbf{g.}}$

It is obvious that the nonholonomic deformation of the Ashtekar connection $%
\ _{\shortmid }A_{\ I}^{I^{\prime }}$ from (\ref{ashtcon}) to the
d--connection $\mathbf{A}_{\ I}^{I^{\prime }}$ (\ref{ndashc}) deforms the
structure of constraints (\ref{constr}), Poisson brackets (\ref{poisson1})
and evolution equations (\ref{eveq}) which makes more sophisticate the
procedure of loop quantization.\footnote{%
from a formal point of view, we get a more cumbersome formulation of the
theory when the Barbero variables are introduced instead of the Ashtekar
ones; nevertheless, this makes the procedure of quantization to be more
physical; here we note that the Ashtekar approach can be generalized for
Riemannian--Cartan spaces \cite{mercuri1} which can be also applied in our
approach even we work with effective torsions induced nonholonomically by
the metric structure, see more details in \cite{vlqgdq2}} Nevertheless, is
spite of a formal sophistication for LQG, the procedure of quantization
become quite simple when there are used DQ methods for the linear connection
$\mathbf{A}_{\ I}^{I^{\prime }}$ because it contains the necessary standard
relation (\ref{torneij}) between the Neijenhuis and torsion tensors, which
in 3--variables (the proof is similar to that for formulas (\ref{ashtdt}) in
Appendix) are written%
\begin{equation}
4\ ^{A}\mathbf{T}_{\ I}^{I^{\prime }}=-\mathbf{\ }^{\mathbf{J}}\mathbf{%
\Omega }_{\ I}^{I^{\prime }},  \label{compconda}
\end{equation}%
where $\ ^{A}\mathbf{T}_{\ I}^{I^{\prime }}$ is the torsion of $\mathbf{A}%
_{\ I}^{I^{\prime }}.$ Following the Fedosov method generalized for almost K%
\"{a}hler and Lagrange--Finsler geometries in Refs. \cite%
{karabeg1,vqg1,vqg2,vqg3,vqg4}, we preserve the variables $_{\shortmid }A_{\
I}^{I^{\prime }}$ and their deformations $\ _{\shortmid }^{A}\mathbf{Z}_{\
I}^{I^{\prime }}$ in the structure of the star product and related quantum
geometric operators.

\subsection{Star products and Fedosov--Ashtekar d--operators in GR}

We introduce the tensor
\begin{equation}
^{L}\mathbf{\Lambda }^{\alpha \beta }\doteqdot \ ^{L}\mathbf{\theta }%
^{\alpha \beta }-i\ \ ^{L}\mathbf{g}^{\alpha \beta },\mbox{ equivalently, }%
\mathbf{\Lambda }^{\alpha ^{\prime }\beta ^{\prime }}\doteqdot \ \mathbf{%
\theta }^{\alpha ^{\prime }\beta ^{\prime }}-i\ \ \mathbf{g}^{\alpha
^{\prime }\beta ^{\prime }},  \label{antsymv}
\end{equation}%
where $\ ^{L}\mathbf{\theta }^{\alpha ^{\prime }\beta }$ is the form (\ref%
{asymstr}) with ''up'' indices and $\ \ ^{L}\mathbf{g}^{\alpha \beta }$ is
the inverse to $\ ^{L}\mathbf{g}_{\alpha \beta }$ (\ref{lfsm}). The local
coordinates on $\mathbf{V}$ are parametrized $u=\{u^{\alpha }\}$ and the
local coordinates on $T_{u}\mathbf{V}$ are labelled $(u,z)=(u^{\alpha
},z^{\beta }),$ where $z^{\beta }$ are fiber coordinates. For a more short
presentation for physicists, we shall omit details on Fedosov's theorems and
their proofs because they are similar to those presented in Refs. \cite%
{fed1,fed2,karabeg1,vqg1,vqg2,vqg3,vqg4} but we shall emphasize certain more
special constructions related to nonholonomic deformations of Ashterkar
variables and related almost K\"{a}hler--Lagrange--Finsler geometric objects.

The formalism of deformation quantization is developed for $C^{\infty }(%
\mathbf{V})[[v]],$ which is the space of formal series in variable $v$ with
coefficients from $C^{\infty }(\mathbf{V})$ on a Poisson manifold $(\mathbf{V%
},\{\cdot ,\cdot \})$ (in this work, we shall consider an almost Poisson
structure generated by the canonical almost symplectic structure in
Lagrange--Finsler and generalized Ashtekar variables). We define an
associative algebra structure on $C^{\infty }(\mathbf{V})[[v]]$ with a $v$%
--linear and $v$--adically continuous star product
\begin{equation}
\ ^{1}f\ast \ ^{2}f=\sum\limits_{r=0}^{\infty }\ _{r}C(\ ^{1}f,\ ^{2}f)\
v^{r},  \label{starp}
\end{equation}%
where $\ _{r}C,r\geq 0,$ are bilinear operators on $C^{\infty }(\mathbf{V})$
with $\ _{0}C(\ ^{1}f,\ ^{2}f)=\ ^{1}f\ ^{2}f$ and $\ _{1}C(\ ^{1}f,\
^{2}f)-\ _{1}C(\ ^{2}f,\ ^{1}f)=i\{\ ^{1}f,\ ^{2}f\};$\ $i$ being the
complex unity. It is possible to introduce a formal Wick product (using
decompositions of type (\ref{starp})),
\begin{equation}
a\circ b\ (z)\doteqdot \exp \left( i\frac{v}{2}\ \mathbf{\Lambda }^{\alpha
\beta }\frac{\partial ^{2}}{\partial z^{\alpha }\partial z_{[1]}^{\beta }}%
\right) a(z)b(z_{[1]})\mid _{z=z_{[1]}},  \label{fpr}
\end{equation}%
for two elements $a$ and $b$ defined by series of type
\begin{equation}
a(v,z)=\sum\limits_{r\geq 0,|\{\alpha \}|\geq 0}\ a_{r,\{\alpha
\}}(u)z^{\{\alpha \}}\ v^{r},  \label{formser}
\end{equation}%
where by $\{\alpha \}$ we label a multi--index. We also consider a formal
Wick algebra $\ \mathbf{W}_{u}\mathbf{=}\ ^{L}\mathbf{W}_{u}$ associated
with the tangent space $T_{u}\mathbf{V}$ enabled with N--connection
splitting induced by an effective $L,$ for any point $u\in \mathbf{V}.$ It
should be noted that the fibre product (\ref{fpr}) can be trivially extended
to the space of $\ ^{L}\mathbf{W}$--valued N--adapted differential forms $\
^{L}\mathcal{W}\otimes \Lambda =\mathcal{W}\otimes \Lambda $ by means of the
usual exterior product of the scalar forms $\Lambda ,$ where $\ \ ^{L}%
\mathcal{W}$ denotes the sheaf of smooth sections of $\ ^{L}\mathbf{W.}$
There are a standard grading on $\Lambda $ denoted $\deg _{a}$ and gradings $%
\deg _{v},\deg _{s},\deg _{a}$ on $\ \ ^{L}\mathcal{W}\otimes \Lambda $
defined on homogeneous elements $v,z^{\alpha },\mathbf{e}^{\alpha }$ as
follows: $\deg _{v}(v)=1,$ $\deg _{s}(z^{\alpha })=1,$ $\deg _{a}(\mathbf{e}%
^{\alpha })=1,$ and all other gradings of the elements $v,z^{\alpha },%
\mathbf{e}^{\alpha }$ are set to zero. In all further constructions $\mathbf{%
e}^{\alpha }$ are N--elongated coframes (\ref{ddif}) or their vierbein
transforms. In this case, the product $\circ $ from (\ref{fpr}) on $\ \ ^{L}%
\mathcal{W}\otimes \mathbf{\Lambda }$ is bigrated. This is written as w.r.t
the grading $Deg=2\deg _{v}+\deg _{s}$ and the grading $\deg _{a}.$

For a 3+1 splitting adapted to an effective Lagrangian $L,$ using triads and
projectors of type
\begin{equation*}
\ \mathbf{q}_{IJ}=\ ^{L}\mathbf{q}_{IJ}=\mathbf{e}_{I}^{\ I^{\prime }}%
\mathbf{e}_{J}^{\ J^{\prime }}\delta _{I^{\prime }J^{\prime }},\ \mathbf{e}%
_{I}^{\ I^{\prime }}\doteqdot \mathbf{e}_{\alpha }^{\ \alpha ^{\prime
}}q_{\alpha ^{\prime }}^{I^{\prime }}q_{I}^{\ \alpha },
\end{equation*}%
we construct $\mathbf{\Lambda }^{IJ}=q_{\alpha }^{I}q_{\beta }^{J}\mathbf{%
\Lambda }^{\alpha \beta }$ stating a formal Wick product for $\ ^{3}\mathbf{%
\Sigma ,}$
\begin{equation*}
\ ^{\Sigma }\circ \ (z)\doteqdot \exp \left( i\frac{v}{2}\ \mathbf{\Lambda }%
^{IJ}\frac{\partial ^{2}}{\partial z^{I}\partial z_{[1]}^{J}}\right)
a(z)b(z_{[1]})\mid _{z=z_{[1]}}.
\end{equation*}%
In this case, we construct a $\ ^{3}\mathbf{\Sigma }$--projection of the
product (\ref{fpr}), denoted ''$\ ^{\Sigma }\circ $'' which is also bigraded
on the space $\ \mathcal{W}\otimes ~^{\Sigma }\mathbf{\Lambda ,}$ where $%
~^{\Sigma }\mathbf{\Lambda }$ is used for $\mathbf{\Lambda }^{IJ}.$

The Ashtekar's d--connection $\mathbf{A}_{\ I}^{I^{\prime }}$ (\ref{ndashc})
is extended to the d--operator
\begin{equation}
\ ^{A}\mathbf{D}\left( a\otimes \lambda \right) \doteqdot \left( \mathbf{e}%
_{\alpha }(a)-u^{\beta }\ \ ^{A}\mathbf{\Gamma _{\alpha \beta }^{\gamma }\ }%
^{z}\mathbf{e}_{\gamma }(a)\right) \otimes (\mathbf{e}^{\alpha }\wedge
\lambda )+a\otimes d\lambda ,  \label{4dahtop}
\end{equation}%
on $\ ^{L}\mathcal{W}\otimes \Lambda ,$ where $^{z}\mathbf{e}_{\alpha }$ is $%
\mathbf{e}_{\alpha }$ redefined in $z$--variables. This canonical almost
symplectic d--connection $\ ^{A}\mathbf{D}$ is a N--adapted $\deg _{a}$%
--graded derivation of the distinguished algebra $\left( \ ^{L}\mathcal{W}%
\otimes \mathbf{\Lambda ,\circ }\right) ,$ in brief, called d--algebra: this
follows from formula (\ref{fpr})). The d--operator (\ref{4dahtop}) projected
on $^{3}\mathbf{\Sigma ,}$ denoted as $^{\Sigma }\mathbf{D,}$ can be written
in an explicit form containing the Ashtekar--Barbero connection and its
nonholonomic deformation (see formulas (\ref{ndashc}) and (\ref{ashtdt})),
\begin{eqnarray*}
&&\ ^{\Sigma }\mathbf{D}\left( a\otimes \mu \right) \doteqdot \left( \mathbf{%
e}^{I^{\prime }}(a)-u^{J}\ \ \mathbf{A}_{~J}^{I^{\prime }~K}\mathbf{\ }^{z}%
\mathbf{e}_{K}(a)\right) \otimes (\mathbf{n}_{I^{\prime }}\wedge \lambda
)+a\otimes d\mu \\
&=&\left( \mathbf{e}^{I}(a)-u^{J}\ (\ \ _{\shortmid }A_{\ J}^{I^{\prime
}~K}+\ _{\shortmid }^{A}\mathbf{Z}_{\ J}^{I^{\prime }~K})\mathbf{\ }^{z}%
\mathbf{e}_{K}(a)\right) \otimes (\mathbf{n}_{I^{\prime }}\wedge \lambda
)+a\otimes d\mu ,
\end{eqnarray*}%
where we took $\mathbf{n}_{I^{\prime }}=\mathbf{n}_{\alpha ^{\prime
}}q_{I}^{\alpha ^{\prime }}$ and $\ \mathbf{A}_{~J}^{I^{\prime }~K}\mathbf{n}%
_{I^{\prime }}=\mathbf{A}_{~J}^{I^{\prime }~}$ and $a\otimes d\mu $ should
be considered on $^{3}\mathbf{\Sigma .}$

The Fedosov distinguished operators (d--operators) $\ \delta $ and $\ \delta
$ $^{-1}$ on $\ \mathcal{W}\otimes \mathbf{\Lambda }$ $\ ($ for $^{L}\delta $
and $\ ^{L}\delta $ $^{-1}$ on $\ ^{L}\mathcal{W}\otimes \mathbf{\Lambda }$
induced by an effective $L,$ we may call them the Fedosov--Lagrange
operators), are defined%
\begin{eqnarray}
\ \delta (a) &=&\ ^{L}\delta (a)=\ \mathbf{e}^{\alpha }\wedge \mathbf{\ }^{z}%
\mathbf{e}_{\alpha }(a),\ \mbox{and\ }\ \   \label{feddop} \\
\delta ^{-1}(a) &=&~^{L}\delta ^{-1}(a)=\left\{
\begin{array}{c}
\frac{i}{p+q}z^{\alpha }\ \ ^{L}\mathbf{e}_{\alpha }(a),\mbox{ if }p+q>0, \\
{\qquad 0},\mbox{ if }p=q=0,%
\end{array}%
\right.   \notag
\end{eqnarray}%
and their projections with respect to $\ $ $^{3}\mathbf{\Sigma ,}$%
\begin{equation}
\ ^{\Sigma }\delta (a)=\ \mathbf{e}^{I}\wedge \mathbf{\ }^{z}\mathbf{e}%
_{I}(a),\ \mbox{and\ }\ ^{\Sigma }\delta ^{-1}(a)=\left\{
\begin{array}{c}
\frac{i}{p+q}z^{I}\ \mathbf{e}_{I}(a),\mbox{ if }p+q>0, \\
{\qquad 0},\mbox{ if }p=q=0,%
\end{array}%
\right. ,  \label{feddop3}
\end{equation}%
where any $a\in \ ^{L}\mathcal{W}\otimes \mathbf{\Lambda }$ is homogeneous
w.r.t. the grading $\deg _{s}$ and $\deg _{a}$ with $\deg _{s}(a)=p$ and $%
\deg _{a}(a)=q.$

The d--operators (\ref{feddop}) define the formula $a=(\ ^{L}\delta \ \
^{L}\delta ^{-1}+\ ^{L}\delta ^{-1}\ \ ^{L}\delta +\sigma )(a),$ where $%
a\longmapsto \sigma (a)$ is the projection on the $(\deg _{s},\deg _{a})$%
--bihomogeneous part of $a$ of degree zero, $\deg _{s}(a)=\deg _{a}(a)=0;$ $%
\delta $ is also a $\deg _{a}$--graded derivation of the d--algebra $\left(
\ \mathcal{W}\otimes \mathbf{\Lambda ,\circ }\right) .$ In a similar form,
using (\ref{feddop3}) on $\left( \ \mathcal{W}\otimes ~^{\Sigma }\mathbf{%
\Lambda ,}~^{\Sigma }\circ \right) ,$ we get the space projection of this
for formula, denoted
\begin{equation*}
^{\Sigma }a=(~^{\Sigma }\delta \ ~^{\Sigma }\delta ^{-1}+~^{\Sigma }\delta
^{-1}\ ~^{\Sigma }\delta +~^{\Sigma }\sigma )(a).
\end{equation*}%
For simplicity, hereafter we shall present only space projected formulas for
indices of geometric objects labelled by $I,J,K,...$ and/or their interior
analogs, ... $I^{\prime },J^{\prime },K^{\prime },...$ if the four
dimensional constructions will not have certain special important properties.

Following a straightforward component calculus similar to that presented in %
\cite{karabeg1,vqg2} (respectively, for holonomic and nonholonomic geometric
configurations), we get the respective torsion and curvature of the Ashtekar
d--operators extended to $\ \mathcal{W}\otimes ~^{\Sigma }\mathbf{\Lambda ,}$%
\begin{equation}
_{z}^{A}\mathcal{T}\ \doteqdot 2~z^{K}\ \ ^{L}\theta _{KI}\mathbf{\ }^{%
\mathbf{J}}\mathbf{\Omega }_{~M}^{I}(u)\ \wedge \mathbf{e}^{M},  \label{at1}
\end{equation}%
for $\ \ ^{L}\theta ^{IJ}=q_{\alpha }^{I}q_{\beta }^{J}\ \ ^{L}\theta
^{\alpha \beta },$ $^{A}\mathbf{T}_{\ M}^{I}=~^{A}\mathbf{T}_{\alpha
M}^{I}(u)\ \mathbf{n}^{\alpha }$ and $4\ ^{A}\mathbf{T}_{\ I}^{I^{\prime }}=-%
\mathbf{\ }^{\mathbf{J}}\mathbf{\Omega }_{\ I}^{I^{\prime }},$ see (\ref%
{compconda}), and
\begin{equation}
\ _{A}^{z}\mathcal{R}\doteqdot \frac{z^{J}z^{K}}{4}\ \ ^{L}\theta _{IJ}\ ^{A}%
\mathbf{F}_{\ KM}^{I}(u)\ \wedge \mathbf{e}^{M},  \label{ac1}
\end{equation}%
where $\ \mathbf{F}_{IJ}^{K}\doteqdot $ $~^{A}\mathbf{R}_{~\alpha
IJ}^{K}(u)\ \mathbf{n}^{\alpha },$ for
\begin{equation*}
\ \mathbf{F}_{IJ}^{K^{\prime }}\doteqdot \mathbf{e}_{I}(\mathbf{A}_{\
J}^{K^{\prime }})-\mathbf{e}_{J}(\ \mathbf{A}_{\ I}^{K^{\prime }})+\epsilon
_{\ I^{\prime }J^{\prime }}^{K^{\prime }}\ \mathbf{A}_{\ I}^{I^{\prime }}\
\mathbf{A}_{\ J}^{J^{\prime }}.
\end{equation*}

By straightforward local computations, we can verify that
\begin{equation}
\left[ \ ^{\Sigma }\mathbf{D},~^{\Sigma }\delta \right] =\frac{i}{v}%
ad_{Wick}(~_{z}^{A}\mathcal{T}\ )\mbox{ and }\ \ ^{\Sigma }\mathbf{D}^{2}=-%
\frac{i}{v}ad_{Wick}(~\ _{z}^{A}\mathcal{R}),  \label{ffedop}
\end{equation}%
where $[\cdot ,\cdot ]$ is the $\deg _{a}$--graded commutator of
endomorphisms of $\ \mathcal{W}\otimes ~^{\Sigma }\mathbf{\Lambda }$ and $%
ad_{Wick}$ is defined via the $\deg _{a}$--graded commutator in $\left(
\mathcal{W}\otimes ~^{\Sigma }\mathbf{\Lambda ,}~^{\Sigma }\circ \right) .$

Having constructed the Fedosov--Ashekar operators, we have defined the main
geometric tools necessary for DG of GR in a form preserving an explicit
relation to variables in LQG.

\section{Deformation Quantization and Ash\-te\-kar d--Con\-nections}

In our previous works \cite{vqg2,vqg3,vqg4}, we proved that formulating a
(pseudo) Riemannian geometry in Lagrange--Finsler variables, we can quantize
the metric, frame and linear connection structures following standard
methods in DQ of almost K\"{a}hler manifolds. Introducing the Ashtekar type
variables in GR, we can follow standard methods of LQG, or (after
corresponding nonholonomic deformations) to apply certain schemes of DQ. The
goal of this section is to re--define the main Fedosov's results \cite%
{fed1,fed2,karabeg1} in generalized Ashtekar variables and show how the
Einstein manifolds can be encoded into the topological structure of
geometrically quantized nonholonomic spaces.

\subsection{Fedosov's approach for Ashtekar d--connections}

Let us consider a (pseudo) Riemannian manifold with the data reformulated in
almost symplectic Ashtekar type variables $^{L}\mathbf{\Lambda }^{\alpha
\beta }$ (\ref{antsymv}) and $\mathbf{A}_{\ I}^{I^{\prime }}$ (\ref{ndashc})
induced by an effective Lagrange structure $L.$

\subsubsection{Definition of flat Fedosov--Ashtekar d--connection}

Any (pseudo) Riemanian metric $\mathbf{g=}\ ^{L}\mathbf{g}_{\alpha \beta }$ (%
\ref{lfsm}), after a 3+1 splitting adapted to a N--connection structure
induced canonically by a regular $L,$ defines a flat normal
Fedosov--Ashtekar d--connec\-ti\-on
\begin{equation}
\ ^{\Sigma }\mathcal{D}\doteqdot -\ \ ^{\Sigma }\delta +\ \ ^{A}\mathcal{D}-%
\frac{i}{v}ad_{Wick}(\ ^{\Sigma }r)  \label{ffadc}
\end{equation}%
satisfying the condition $\ ^{\Sigma }\mathcal{D}^{2}=0,$ where the unique
element $\ ^{\Sigma }r\in $ $\ ^{L}\mathcal{W}\otimes ~^{\Sigma }\mathbf{%
\Lambda ,}$ $\deg _{a}(\ ^{\Sigma }r)=1,$ $\ ^{\Sigma }\delta ^{-1}\
^{\Sigma }r=0,$ solves the equation
\begin{equation}
\ \ \ ^{\Sigma }\delta r=~_{\Sigma }^{A}\mathcal{T}\ +~^{A}\mathcal{F}+\ \
^{A}\mathcal{D}\ ^{\Sigma }r-\frac{i}{v}\ ^{\Sigma }r\circ \ ^{\Sigma }r,
\label{eqffadc}
\end{equation}%
where $~_{\Sigma }^{A}\mathcal{T}\ $\ and $~^{A}\mathcal{F}$ on hypersurface
$\ ^{3}\mathbf{\Sigma }$ are respectively the 2--forms of torsion and
curvature of $\ ^{A}\mathcal{D}=\{\mathbf{A}_{\ I}^{I^{\prime }}\}.$ The
element $\ ^{\Sigma }\delta r$ is computed recursively with respect to the
total degree $Deg$ as follows:%
\begin{eqnarray*}
\ ^{\Sigma }r^{(0)} &=&\ ^{\Sigma }r^{(1)}=0,\ ^{\Sigma }r^{(2)}=\ ^{\Sigma
}\delta ^{-1}~_{\Sigma }^{A}\mathcal{T}, \\
\ ^{\Sigma }r^{(3)} &=&\ ^{\Sigma }\delta ^{-1}\left( ~^{A}\mathcal{F}+\
^{\Sigma }\mathcal{D}\ ^{\Sigma }r^{(2)}-\frac{i}{v}\ ^{\Sigma }r^{(2)}\circ
\ ^{\Sigma }r^{(2)}\right) , \\
\ ^{\Sigma }r^{(k+3)} &=&\ ^{\Sigma }\delta ^{-1}\left( \ ^{\Sigma }\mathcal{%
D}\ ^{\Sigma }r^{(k+2)}-\frac{i}{v}\sum\limits_{l=0}^{k}\ ^{\Sigma
}r^{(l+2)}\circ \ ^{\Sigma }r^{(l+2)}\right) ,k\geq 1,
\end{eqnarray*}%
where by $a^{(k)}$ we denoted the $Deg$--homogeneous component of degree $k$
of an element $a\in $ $\ \ \mathcal{W}\otimes ~^{\Sigma }\mathbf{\Lambda }.$
The proof of formulas \ (\ref{ffadc})\ and (\ref{eqffadc})\ consists from
straightforward verifications of the property $\ ^{\Sigma }\mathcal{D}^{2}=0$
using for $\ ^{\Sigma }r$ a formal series of type (\ref{formser}).

The above--presented formulas are written in a four dimensional form for the
d--connection $\ ^{A}\mathbf{\Gamma }_{\beta \gamma }^{\alpha }\mathbf{(\
\mathbf{g})}$ (\ref{defashtd}) and Fedosov--Lagrange operators $\ \delta $
and $\ \delta $ $^{-1}$ (\ref{feddop}). From a formal point of view, we have
to omit the label ''$\Sigma "$ and rewrite the formulas for the torsion $^{A}%
\mathcal{T}$ \ (\ref{torneij}) and curvature $^{A}\mathcal{R},$ when $^{A}%
\mathcal{R}_{\ \gamma }^{\tau }=~^{A}\mathbf{R}_{\ \gamma \alpha \beta
}^{\tau }\ \mathbf{e}^{\alpha }\wedge \ \mathbf{e}^{\beta };$ for
coefficients, see similar formulas (\ref{curv}), but for this case
considered for the d--connection $\ ^{A}\mathbf{\Gamma }_{\beta \gamma
}^{\alpha }.$

There is a flat connection
\begin{equation}
\ \mathcal{D}\doteqdot -\ \ \delta +\ \ ^{A}\mathcal{D}-\frac{i}{v}%
ad_{Wick}(\ r)  \label{ffdc4}
\end{equation}%
satisfying the condition $\ \mathcal{D}^{2}=0,$ where the unique element $%
r\in $ $\ \mathcal{W}\otimes ~^{\Sigma }\mathbf{\Lambda ,}$ $\deg _{a}(r)=1,$
$\ \delta ^{-1}\ r=0,$ solves the equation
\begin{equation}
\delta r=~^{A}\mathcal{T}\ +~^{A}\mathcal{R}+\ \ ^{A}\mathcal{D}\ r-\frac{i}{%
v}\ r\circ \ r,  \label{eqffadc4}
\end{equation}%
where $~$the element $\ \delta r$ can be computed recursively with respect
to the total degree $Deg$ as follows:%
\begin{eqnarray*}
\ r^{(0)} &=&\ r^{(1)}=0,\ r^{(2)}=\ \delta ^{-1}~^{A}\mathcal{T}, \\
\ r^{(3)} &=&\ \delta ^{-1}\left( ~^{A}\mathcal{R}+\ \mathcal{D}\ r^{(2)}-%
\frac{i}{v}\ r^{(2)}\circ \ r^{(2)}\right) , \\
\ r^{(k+3)} &=&\ \delta ^{-1}\left( \ \mathcal{D}\ r^{(k+2)}-\frac{i}{v}%
\sum\limits_{l=0}^{k}\ r^{(l+2)}\circ \ r^{(l+2)}\right) ,k\geq 1.
\end{eqnarray*}%
We note that the formulas \ (\ref{ffdc4})\ and (\ref{eqffadc4}) are
ismorphic to similar ones considered in Theorem 3.1 from Ref. \cite{vqg3},
but re--adapted in this work for a 3+1 splitting. Their space three
dimensional projections result in (\ref{ffadc})\ and (\ref{eqffadc})\
containing a nonholonomic and quantum deformation of the Ashtekar--Barbero
variables.

\subsubsection{The star--product induced by the Fedosov--Ashtekar
d--con\-nection}

We present the definition of the star--product for three and four
dimensional hypersurfaces with the d--connections canonically related\ both
to the Ashtekar variables and Finsler--Lagrange ones.

\paragraph{ Three dimensional hypersurface constructions:\ }

The procedure of deformation quantization is strongly related to the
definition of a star--product which in our approach is computed canonically
because the Ashtekar d--connection $\mathbf{A}_{\ I}^{I^{\prime }}$ (\ref%
{ndashc}) is a N--adapted affine and/or almost symplectic connection like
those considered in \cite{karabeg1,vqg2,vqg3,vqg4}. A star--product on the
almost K\"{a}hler model of a (pseudo) Riemannian space in Lagrange--Finsler
variables enabled with a canonicaly nonholonomically deformed
Ash\-tekar--Barbero connection is defined on $C^{\infty }(\mathbf{V})[[v]]$
by formula
\begin{equation}
\ ^{1}f~\ ^{\Sigma }\ast \ ^{2}f\doteqdot \ ^{\Sigma }\sigma (\ ^{\Sigma
}\tau (\ ^{1}f))\circ \ ^{\Sigma }\sigma (\ ^{\Sigma }\tau (\ ^{2}f)),
\label{asp3}
\end{equation}%
where the projection $^{\Sigma }\sigma :\ \ \mathcal{W}_{\mathbf{A}%
}\rightarrow C^{\infty }(\mathbf{V})[[v]]$ onto the part of $\deg _{s}$%
--degree zero is a bijection and the inverse map $\ ^{\Sigma }\tau
:C^{\infty }(\mathbf{V})[[v]]\rightarrow \ \ \mathcal{W}_{\mathbf{A}}$ can
be calculated recursively w.r..t the total degree $Deg,$%
\begin{eqnarray*}
\ ^{\Sigma }\tau (f)^{(0)} &=&f\mbox{\ and, for \ }k\geq 0, \\
\ ^{\Sigma }\tau (f)^{(k+1)} &=&^{\Sigma }\delta ^{-1}[\ ^{A}\mathbf{D}\
^{\Sigma }\tau (f)^{(k)}-\frac{i}{v}\sum\limits_{l=0}^{k}ad_{Wick}(\
^{\Sigma }r^{(l+2)})(^{\Sigma }\tau (f)^{(k-l)})].
\end{eqnarray*}

We denote by $\ ^{f}\xi $ the Hamiltonian vector field corresponding to a
function $f\in C^{\infty }(\mathbf{V})$ on space $(\mathbf{V},\ \theta )$
and consider the antisymmetric part%
\begin{equation*}
\ ^{-}C(\ ^{1}f,\ ^{2}f)\ \doteqdot \frac{1}{2}\left( C(\ ^{1}f,\ ^{2}f)-C(\
^{2}f,\ ^{1}f)\right)
\end{equation*}%
of bilinear operator $C(\ ^{1}f,\ ^{2}f).$ We say that a star--product (\ref%
{starp}) is normalized if $\ _{1}C(\ ^{1}f,\ ^{2}f)=\frac{i}{2}\ ^{\Sigma
}\{\ ^{1}f,\ ^{2}f\},$ where $\ ^{\Sigma }\{\cdot ,\cdot \}$ is the Poisson
bracket defined with respect to the space hypersurfice $\ ^{3}\mathbf{\Sigma
.}$ For the so--called normalized $~^{\Sigma }\ast ,$ the bilinear operator $%
\ _{2}^{-}C$ defines a de Rham--Chevalley 2--cocycle, when there is a unique
closed 2--form $\ ~^{\Sigma }\varkappa $ such that%
\begin{equation}
\ _{2}C(\ ^{1}f,\ ^{2}f)=\frac{1}{2}\ \ ~^{\Sigma }\varkappa (\ ^{f_{1}}\xi
,\ ^{f_{2}}\xi )  \label{c2}
\end{equation}%
for all $\ ^{1}f,\ ^{2}f\in C^{\infty }(\mathbf{V}).$ This is used to
introduce $c_{0}(\ ^{\Sigma }\ast )\doteqdot \lbrack \ \ ~^{\Sigma
}\varkappa ]$ as the equivalence class.

A straightforward computation of $\ _{2}C$ from (\ref{c2}) and previous
formulas prove that there is a unique 2--form defined by the Ashtekar
d--connection $\mathbf{A}_{\ I}^{I^{\prime }}$ (\ref{ndashc}),%
\begin{equation}
\ ^{\Sigma }\varkappa =-\frac{i}{8}\mathbf{J}_{K^{\prime }}^{\ I}~\mathbf{F}%
_{IJ}^{K^{\prime }}\wedge \mathbf{e}^{J}-\frac{i}{6}d\left( \mathbf{J}%
_{I^{\prime }}^{\ I}~^{A}\mathbf{T}_{\ I}^{I^{\prime }}\right) ,  \label{2f3}
\end{equation}%
where $\mathbf{F}_{IJ}^{K^{\prime }}\doteqdot $ $~^{A}\mathbf{R}_{~\alpha
IJ}^{K^{\prime }}(u)\ \mathbf{n}^{\alpha }$ and $\mathbf{J}_{I^{\prime }}^{\
I}=\mathbf{J}_{\alpha ^{\prime }}^{\ \beta }q_{\alpha }^{I}q_{I^{\prime
}}^{\ \alpha ^{\prime }}.$

Let us define another canonical class $\ ~^{\Sigma }\varepsilon ,$ for $\
^{N}T\ ^{3}\Sigma =h\ ^{3}\Sigma \oplus v\ ^{3}\Sigma $ induced by $\ ^{N}T%
\mathbf{V}=h\mathbf{V}\oplus v\mathbf{V}$ (\ref{whitney}), where the left
label points that the tangent bundle is split nonholonomically by the
canonical N--connection structure $\mathbf{N}.$ We perform a distinguished
complexification of such second order tangent bundles in the form $T_{%
\mathbb{C}}\left( \ ^{N}T\mathbf{V}\right) =T_{\mathbb{C}}\left( h\mathbf{V}%
\right) \oplus T_{\mathbb{C}}\left( v\mathbf{V}\right) ,$ then consider $T_{%
\mathbb{C}}\left( \ ^{N}T\ ^{3}\Sigma \right) =T_{\mathbb{C}}\left( h\
^{3}\Sigma \right) \oplus T_{\mathbb{C}}\left( v\ ^{3}\Sigma \right) .$ We
introduce $~^{\Sigma }\varepsilon $ as the first Chern class of the
distributions $T_{\mathbb{C}}^{\prime }\left( \ ^{N}T~^{3}\Sigma \right) =T_{%
\mathbb{C}}^{\prime }\left( h~^{3}\Sigma \right) \oplus T_{\mathbb{C}%
}^{\prime }\left( v~^{3}\Sigma \right) $ $\ $induced by $T_{\mathbb{C}%
}^{\prime }\left( \ ^{N}T\mathbf{V}\right) =T_{\mathbb{C}}^{\prime }\left( h%
\mathbf{V}\right) \oplus T_{\mathbb{C}}^{\prime }\left( v\mathbf{V}\right) $
of couples of vectors of type $(1,0)$ both for the h-- and v--parts. In
explicit form, we calculate $~^{\Sigma }\varepsilon $ using the
d--connection $\ ^{A}\mathcal{D}$ (\ref{ndashc}) and the h- and
v--projections $h\Pi =\frac{1}{2}(Id_{h}-iJ_{h})$ and $v\Pi =\frac{1}{2}%
(Id_{v}-iJ_{v}),$ with respective restrictions on $\ ^{3}\Sigma ,$ $%
h~^{\Sigma }\Pi $ and $v~^{\Sigma }\Pi ,$ where $Id_{h}$ and $Id_{v}$ are
respective identity operators and $J_{h}$ and $J_{v}$ are almost complex
operators, which are projection operators onto corresponding $(1,0)$%
--subspaces. We consider the the matrix $\left( h\Pi ,v\Pi \right) \ ^{A}%
\mathcal{R}\left( h\Pi ,v\Pi \right) ^{T},$ where $(...)^{T}$ means
transposition, as the curvature matrix of the N--adapted restriction of $\ $%
d--connection$\ \ ^{A}\mathcal{D}$ to $T_{\mathbb{C}}^{\prime }\left( \ ^{N}T%
\mathbf{V}\right) .$ The restriction of this matrix on $\ ^{3}\Sigma $ gives
$\left( h~^{\Sigma }\Pi ,v~^{\Sigma }\Pi \right) \ ~^{A}\mathcal{F}\left(
h~^{\Sigma }\Pi ,v~^{\Sigma }\Pi \right) ^{T}.$ This allows us to compute
the closed Chern--Weyl form
\begin{eqnarray}
\ ~^{\Sigma }\gamma &=&-iTr\left[ \left( h~^{\Sigma }\Pi ,v~^{\Sigma }\Pi
\right) ~^{A}\mathcal{F}\left( h~^{\Sigma }\Pi ,v~^{\Sigma }\Pi \right) ^{T}%
\right]  \label{aux4} \\
&=&-iTr\left[ \left( h~^{\Sigma }\Pi ,v~^{\Sigma }\Pi \right) ~^{A}\mathcal{F%
}\right] = -\frac{1}{4}\mathbf{J}_{K^{\prime }}^{\ I}~\mathbf{F}%
_{IJ}^{K^{\prime }}\wedge \mathbf{e}^{J}  \notag
\end{eqnarray}%
and define the canonical class is $~^{\Sigma }\varepsilon \doteqdot \lbrack
\ ~^{\Sigma }\gamma ].$

We conclude that the zero--degree cohomology coefficient for the almost K%
\"{a}hler model of a (pseudo) Riemannian space with Ashtekar d--connection
is $c_{0}(\ ^{\Sigma }\ast )=-(1/2i)\ ~^{\Sigma }\varepsilon .$

\paragraph{Four dimensional constructions:}

In this case, we work directly with the d--connection $\ ^{A}\mathbf{D=\{}\
^{A}\mathbf{\Gamma }_{\beta \gamma }^{\alpha }\}$ (\ref{defashtd}). For
simplicity, we omit the proofs with are similar to the three dimensional
case and present only the most important formulas. The formal formula for
the star--product is
\begin{equation}
\ ^{1}f~\ \ast \ ^{2}f\doteqdot \ \sigma (\ \tau (\ ^{1}f))\circ \ \sigma (\
\tau (\ ^{2}f)),  \label{asp3a}
\end{equation}%
where the projection $\sigma :\ \ \mathcal{W}_{^{A}\mathbf{\Gamma }%
}\rightarrow C^{\infty }(\mathbf{V})[[v]]$ onto the part of $\deg _{s}$%
--degree zero is a bijection and the inverse map $\tau :C^{\infty }(\mathbf{V%
})[[v]]\rightarrow \ \mathcal{W}_{^{A}\mathbf{\Gamma }}$ can be calculated
recursively w.r.t. the total degree $Deg,$%
\begin{eqnarray*}
\ \tau (f)^{(0)} &=&f\mbox{\ and, for \ }k\geq 0, \\
\ \tau (f)^{(k+1)} &=&\ \delta ^{-1}[\ ^{A}\mathbf{D}\ \tau (f)^{(k)}-\frac{i%
}{v}\sum\limits_{l=0}^{k}ad_{Wick}(\ r^{(l+2)})(\ \tau (f)^{(k-l)})].
\end{eqnarray*}

The bilinear operator $\ _{2}^{-}C$ defining the de Rham--Chevalley
2--cocycle for the normalized $\ast $ is given by a unique closed 2--form $%
\varkappa $ such that%
\begin{equation*}
\ _{2}C(\ ^{1}f,\ ^{2}f)=\frac{1}{2}\ \ \varkappa (\ ^{f_{1}}\xi ,\
^{f_{2}}\xi ),
\end{equation*}%
for all $\ ^{1}f,\ ^{2}f\in C^{\infty }(\mathbf{V}),$ which is used to
introduce $c_{0}(\ast )\doteqdot \lbrack \varkappa ]$ as the equivalence
class with

\begin{equation}
\varkappa =-\frac{i}{8}\mathbf{J}_{\tau }^{\ \alpha ^{\prime }}\ ^{A}%
\mathcal{R}_{\ \alpha ^{\prime }}^{\tau }-\frac{i}{6}d\left( \mathbf{J}%
_{\tau }^{\ \alpha ^{\prime }}~^{A}\mathbf{T}_{\ \alpha ^{\prime }\beta
}^{\tau }\ \mathbf{e}^{\beta }\right) .  \label{2f3a}
\end{equation}%
where the coefficients of the curvature and torsion 2--forms of the normal
d--connection 1--form are given respectively by formulas (\ref{cform}) and (%
\ref{tform}).

The closed Chern--Weyl form is computed
\begin{eqnarray}
\gamma &=&-iTr\left[ \left( h\Pi ,v\Pi \right) ~^{A}\mathcal{R}\left( h\Pi
,v\Pi \right) ^{T}\right]  \label{aux4a} \\
&=&-iTr\left[ \left( h\Pi ,v\Pi \right) ~^{A}\mathcal{R}\right] = -\frac{1}{4%
}\mathbf{J}_{\tau }^{\ \alpha ^{\prime }} \ ^A\mathcal{R}_{\ \alpha^{\prime
}}^{\tau },  \notag
\end{eqnarray}%
which defines the canonical class is $~\varepsilon \doteqdot \lbrack \
\gamma ]$ and the zero--degree cohomology coefficient for the almost K\"{a}%
hler model of a (pseudo) Riemannian space with d--connection $\ ^{A}\mathbf{%
D,}$ computed $c_{0}(\ast )=-(1/2i)\varepsilon .$

Formulas (\ref{asp3a}), (\ref{2f3a}) and (\ref{aux4a}) \ were computed
following the Karabegov and Schlichenmaier approach \cite{karabeg1} for the
canononical d--connection and normal d--connection in Lagrange--Finsler
geometry and Einstein spaces enabled with nonholonomic distiributions
modelled on nonholonomic manifolds or tangent bundles, see respecively \cite%
{vqg1,vqg2,vqg3,vqg4}. Working with the d--connection $\ ^{A}\mathbf{D,}$
all constructions are generated in adapted forms both to the nonholonomic
2+2 and 3+1 splitting which allows us to obtain on corresponding three
dimensional hypersurfaces the respective formulas (\ref{asp3}), (\ref{2f3})
and (\ref{aux4}) containing nonholonomic deformations of the
Ashtekar--Barbero connection.

\subsection{The zero--degree cohomology for Einstein spaces in
Ash\-te\-kar--Lagrange variables}

The 3+1 splitting formalism encode the Einstein equations in an equivalent
'quantum' form as the Wheeler De Witt equations. In new canonical (Ashtekar
and modifications) variables, for LQG, one has to consider the so--called
''master constraint programme''. So, for certain canonical and loop
quantization models, the gravitational field equations are considered as the
starting point for definition of the quantization formalism.

In DQ, there were elaborated quantization schemes when metric, vielbein and
connection fields are not obligatory subjected to satisfy certain field
equations and/or derived by a variational procedure. For instance, such
geometric and/or BRST quantization approaches were proposed in Ref. \cite%
{lyakh1,lyakh2}. This allows us to apply DQ methods to generalized theories
of gravity, matter field interactions and nonlinear mechanical systems.

We proved in the previous sections that under certain well defined
conditions the scheme of LQG is equivalent to that from DG in a
nonholonomically modified Fedosov approach. Even in such generalized
Ashtekar--Lagrange variables (with $\mathbf{A}_{\ I}^{I^{\prime }}$ (\ref%
{ndashc}) and $\ ^{A}\mathbf{\Gamma }_{\beta \gamma }^{\alpha }$ (\ref%
{defashtd})) the Poisson structure (\ref{poisson1}) and related evolution
equations (\ref{eveq}) are nonholonomically deformed and we ''loose'' a
direct connection to the Einstein equations, the scheme of DQ gives us the
possibility to consider nonholonomic quantum deformations of the
gravitational field equations \cite{vqg3,vqg4}. Here, we analyze in brief
the problem of encoding the Einstein equations into the formalism of
nonholonomic DQ for the d--connection $\ ^{A}\mathbf{\Gamma }_{\beta \gamma
}^{\alpha }$ (using formulas (\ref{ashtaux}) and (\ref{ashtdt}), we can
compute all values corresponding to $\mathbf{A}_{\ I}^{I^{\prime }}).$

Introducing the formulas (\ref{veinst1}) into formulas (\ref{2f3a}) and (\ref%
{aux4a}), we obtain that the zero--degree cohomology coefficient $c_{0}(\ast
)$ for the almost K\"{a}hler model of an Einstein space defined by a
d--tensor $\mathbf{g=}\ ^{L}\mathbf{g}_{\alpha \beta }$ (\ref{lfsm}) as a
solution of \ the vacuum Einstein equations) is $c_{0}(\ast
)=-(1/2i)\varepsilon ,$ for $\varepsilon \doteqdot \lbrack \ \gamma ],$
where
\begin{equation}
\gamma =\frac{1}{4}\mathbf{J}_{\tau \alpha }^{\ }\left( -\lambda \mathbf{e}%
^{\tau }\wedge \ \mathbf{e}^{\alpha }+\widehat{\mathcal{Z}}^{\tau \alpha
}\right) .  \label{cwform}
\end{equation}
It should be noted that for $\lambda =0$ the 2--form $\widehat{\mathcal{Z}}%
^{\tau \alpha }$ is defined by the deformation d--tensor from the
Levi--Civita connection to the normal d--connection (\ref{cdeft}), see
formulas (\ref{cdeftc}). We conclude that $c_{0}(\ast )$ encodes the vacuum
Einstein configurations, in general, with nontrivial cosmological constant $%
\lambda$ and their quantum deformations.

Multiplying $\mathbf{e}^{\alpha }\wedge $ with (\ref{cwform}) written in
Lagrange--Finsler variables with a further 3+1 decomposition and introducing
the almost complex operator $\mathbf{J}_{\beta \gamma },$ we get the almost
symplectic form of the Einstein equations (\ref{veinst1}). In
Ashtekar--Lagrange variables, the quantum field equations corresponding to
the Einstein gravitations in general relativity are
\begin{equation}
\mathbf{e}^{\alpha }\wedge \gamma =\epsilon ^{\alpha \beta \gamma \tau }2\pi
G\mathbf{J}_{\beta \gamma }\widehat{\mathcal{T}}_{\ \tau }\ -\frac{\lambda }{%
4}\mathbf{J}_{\beta \gamma }\mathbf{e}^{\alpha }\wedge \mathbf{e}^{\beta
}\wedge \ \mathbf{e}^{\gamma }.  \label{aseq}
\end{equation}

We emphasize that in the vacuum case, when $\lambda =0,$ the 2--form $\gamma
$ (\ref{cwform}) from (\ref{aseq}) is not zero but defined by $\widehat{%
\mathcal{T}}_{\ \tau }=\ ^{Z}\widehat{\mathcal{T}}_{\ \tau }.$ Finally, we
note that it is possible an explicit computation of $\gamma $ for nontrivial
matter fields following combined LQG and DQ approaches to models with
interacting gravitational and matter fields geometrized in terms of an
almost K\"{a}hler model defined for spinor and fiber bundles on spacetime %
\cite{vlqgdq2}.

\section{Conclusions}

There are four conventional communities of researches working in modern
quantum gravity, separated by different philosophies and purposes in
physics, mathematical 'language' and ways of using mathematical formalism.
Two communities related to string/ M--theory (SMT) and perturbative quantum
gravity (PQG) have a number of common points, for instance, with the
background field method but they propose different approaches and claim to
have a different status in theoretical particle physics. The first one was
built as a theory of unification of interactions and the second one is still
considered to be undefined because of non--renormalizibility and a number of
conceptual and technical problems. Here, one should be emphasized that PQG
can be more simply and directly standard theories of physics and
experimental data. Researches from SMT and loop quantum gravity (LQG, the
involved persons consist the third community) have a common goal, a
consistent theory of quantum gravity, but both theories are with very
different underlying hypothesis, mathematical tools, a lot of different
complementary tasks and yet unresolved theoretical problems and lack of
experimental verification. It is still supposed that LQG might become a part
of SMT but this will be not the case if, for instance, the supersymmetry is
not necessary for a consistent quantization of gravity.

In LQG a lot of serious work has been done and considerable progress was
achieved. Some researches still point to possible problems with the
semi--classical limit, compatibility with the perturbative approach and
spacetime covariance and diffeomorphysm invariance \cite{nicolai}.
Nevertheless, it is considered that the constructions in LQG are physically
well motivated and (sometimes) mathematically unique which makes the
approach less ambiguous \cite{thiem} but still less familiar for particle
physics researches.

During last three decades, it was also an intensive work on geometric and
deformation quantization (DQ, we classify the authors publishing in this
direction to belong to the forth community in quantum gravity), see
reflections \cite{sternh1} on philosophy of DQ. The Fedosov quantization %
\cite{fed1,fed2} was generalized for Poisson manifolds \cite{konts1} and
almost K\"{a}hler spaces \cite{karabeg1}. For a long time, it was considered
that DQ might not have direct applications to the general relativity (GR)
theory because torsion and almost symplectic connections play a crucial role
in DQ, which was considered to be not related to the Levi--Civita connection
in (pseudo) Riemannian geometry.

We proved that the classical Einstein theory can be re--formulated
equivalently as an almost K\"{a}hler geometry if the (pseudo) Riemannian
spacetime is enabled with a formal (2+2)--dimensional nonholonomic
distribution \cite{vqg1,vqg2,vqg3,vqg4}. For such a distribution inducing an
effective nonlinear connection, a (pseudo) Riemannian metric defines not
only the preferred Levi--Civita linear connection (which, by definition, is
metric compatible and torsionless) but also an infinite number of metric
compatible linear connections with nontrivial torsions. There is a canonical
connection when the torsion is completely and uniquely induced by the metric
coefficients in a form necessary for DQ of almost complex/ symplectic
structures. So, we can quantize GR by deforming it in a unique form to a
corresponding model of nonholonomic almost K\"{a}hler geometry (all data for
deformations of geometric objects, their values on the 'primary' (pseudo)
Riemannian spacetime and 'target' almost symplectic nonholonomic manifolds
being defined by a 'primary' metric structure). Such generalized transforms
are described not only by frame and coordinate maps but also by deformations
(equivalently, distorsions) of connections and fundamental tensor/spinor
objects. They result in effective Riemann--Cartan (or Einstein--Cartan)
spaces when the torsion is completely induced by certain generic
'off--diagonal' coefficients of a metric related to the 2+2 nonholonomic
splitting.

A very important geometric techniques for DQ of GR was taken from Finsler
geometry and its generalization as the Lagrange geometry \cite{ma} (we note
the formalism of nonlinear connections, N--connections, the geometry of
nonholonomic manifolds and N--adapted linear connections and geometric
objects \cite{vr1,vr1a,vr2,bejf,vsgg}). In our approach, we emphasized that
various types of Lagrange--Finsler variables can be introduced formally also
on (pseudo) Riemannian manifolds, which is very useful for parametrizations
of certain very general classes of generic off--diagonal solutions of the
Einstein equations in GR, string gravity and gauge (non) commutative
gravity, see original results and reviews \cite{vrfg,vsgg,vncg,vijgmmp}.

It should be emphasized here that some groups of mathematicians and
physicists use in their investigations on generalized gravity models and
applications of Finsler geometry the so--called Cartan connection \cite{bcs}
which is a torsionless but metric noncompatible generalization of the
Levi--Civita connection. Nonmetricity, in such cases, makes more
sophisticate and almost impossible (without a metric compatible background)
the definition of a number of very important physical geometric objects and
concepts (like spinors, twistors and conservation laws) and formulation of
noncommutative generalizations of Finsler geometry, which also exists in
Lagrange--Finsler DQ. For the purposes considered in this work, elaboration
of an unified approach to LQG and DQ of gravity and further applications on
standard physics, a nonmetricity field would induce a number of very
difficult problems for quantization of models with locally anisotropic
spaces (see detailed discussions and reviews of results in \cite{vrfg,vsgg}%
). In our works on Lagrange--Finsler and nonholonomic manifolds geometry, we
follow a more 'conservative' and 'pragmatic' point of view when locally
anisotropic configurations are modelled by nonholonomic distributions on
(pseudo) Riemannian and Riemann--Cartan spaces and the condition of metric
compatibility of linear connections is positively satisfied. This is the
case for the so--called Cartan connection in Finsler geometry \cite{cart}
and various metric compatible modifications and generalizations of nonlinear
and linear connections \cite{ma,bej,vrfg,vsgg,vncg,vijgmmp} which can be
related to certain classes of nonholonomic almost symplectic/K\"{a}hler
structures and presents a substantial interest for application of DQ methods.

For various purposes of DQ of GR, to introduce and work with formal
(effective) Lagran\-ge--Finsler variables is very important because this
allows us to construct in explicit form some canonical symplectic forms and
linear connections completely defined by a metric structure (which may be,
or not, a solution of the Einstein equations). If such geometric and
physical objects are introduced on (co) tangent bundles, we perform DQ
models for general nonlinear mechanical models encoded as Lagrange--Finsler
(Hamilton--Cartan) geometries; here we note that, for instance, the Finsler
geometry consists a particular (homogeneous) case of Lagrange geometry. We
can consider an inverse situation when fixing a convenient formal (regular
Lagrange) generating function on a (pseudo) Riemannian manifold (in a
particular case, we take an Einstein manifold), we model effectively certain
gravitational processes as analogous mechanics systems. This allows us also
to model quantum GR effects by certain quantized nonlinear mechanics
Lagrange/ Hamilton models following well defined methods of nonholonomic
quantum geometry.

The aim of this article was to give a self--contained and comparative
analysis of two existing general approaches to quantization of GR, the LQG
and DQ, and to construct a "bridge" between these two quantum gravity
theories. In spite of their general difference on methods and philosophy,
both approaches contain certain common principles and ideas. Both
quantizations are based on the idea of spacetime splitting preserving the
general covariance and diffeomorphism invariance: a 3+1 splitting is used in
LQG and a 2+2 nonholonomic splitting is used in DQ. In both cases, certain
new variables are necessary to perform the procedure of quantization: for
LQG, one uses the Ashtekar--Barbero variables, which simplifies the
structure of constraints, and, for DG, one considers the Lagrange--Finsler
variables, which allows us to define certain canonical almost symplectic
fundamental geometric objects.

There are also substantial differences between the above--mentioned two
approaches. For instance, in LQG, the procedure of quantization is strongly
determined by the Wheeler -- de Witt equations (which are equivalent to the
Einstein equations, define an effective Hamiltonian for gravity with a 3+1
splitting and conclude in a corresponding algebras of constraints, for
different classes of new variables). It is also very important the
so--called Master Constraint Programme \cite{thiem} allowing to define a
quantum solution for constraints and quantize the theory following the Dirac
method. The procedure of DQ can be performed for any almost symplectic /
Poisson structure enabled with a necessary type linear connection when the
torsion is defined by the Neijenhuis tensor. In this case, the condition to
find in explicit form a quantum variant of Einstein equations and to solve
certain classes of constraints is not so important, all data being encoded
into the nonholonomic configuration of theory and the zero--degree
cohomology coefficient (where a nonholonomically deformed version of the
Einstein equations is contained). We conclude here that LQG may be more
convenient for establishing a self--consistent semi--classical limit when
the form of classical Einstein equations is well known. The de--quantization
procedure in DQ and (in general) the formalism of DQ are crucial if certain
classical models (their fundamental geometric objects, field equations and
conservation laws) would be substantially modified under quantization (for
instance, the BRST formalism together with DQ consider methods of
non--Lagrange theories and other various exotic quantum models \cite%
{lyakh1,lyakh2}).

\begin{figure}[htbp]
\begin{center}
\begin{picture}(360,500)
\thinlines
\put(190,480){\oval(160,35)}
\put(114,478){\makebox{(pseudo) Riemannian metric ${\bf g}$}}
\put(80,420){\makebox{\it 3+1 splitting}}
\put(240,420){\makebox{\it 2+2 nonholonomic splitting}}
\put(0,350){\framebox(120,40){ADM formalism}}
\put(255,350){\framebox(160,40)
{$\begin{array}{c}
 \mbox{Lagrange--Finsler variables,}\\
 \mbox{Nonlinear Connection formalism}
 \end{array}$
}}
\put(30,270){\framebox(110,40)
{$\begin{array}{c}
 \mbox{Ashtekar--Barbero}\\
 \mbox{variables}
 \end{array}$
 }}
\put(240,270){\framebox(135,40)
{$\begin{array}{c}
 \mbox{Almost K\"{a}hler}\\
 \mbox{nonholonomic variables}
 \end{array}$
 }}
 \put(0,200){\framebox(110,40){LQG}}
   \put(270,200){\framebox(135,40)
{Fedosov--Lagrange DQ}}
\put(110,155){\dashbox(160,40)
{$\begin{array}{c}
 \mbox{Ashtekar \& Almost Symplectic}\\
 \mbox{canonical d--connections}
 \end{array}$
}}
\put(75,100){\dashbox(240,40)
{$\begin{array}{c}
 \mbox{Master Constraint Program \& Zero Degree }\\
 \mbox{Cohomology, Quantum Einstein equations}
 \end{array}$
 }}
\put(288,350){\shortstack[r]{ \vector(0,-1){40}}}
\put(92,350){\shortstack[r]{ \vector(0,-1){40}}}
\put(288,270){\shortstack[r]{ \vector(0,-1){30}}}
\put(92,270){\shortstack[r]{ \vector(0,-1){30}}}
\put(288,197){\shortstack[r]{ \vector(-1,-1){15}}}
\put(92,197){\shortstack[r]{ \vector(1,-1){15}}}
\put(190,152){\shortstack[r]{ \vector(0,-1){12}}}
\put(190,100){\shortstack[r]{ \vector(0,-1){50}}}
\put(97,10){\framebox(170,40)
{$\begin{array}{c}
 \mbox{Generalized Quantum Theories}\\
 \mbox{and Applications}
 \end{array}$
}}
\put(190,460){\shortstack[r]{\vector(1,-1){70}}}
\put(190,460){\shortstack[r]{\vector(-1,-1){70}}}


\end{picture}
\end{center}
\caption{\textbf{LQG and DQ}}
\label{fig1}
\end{figure}
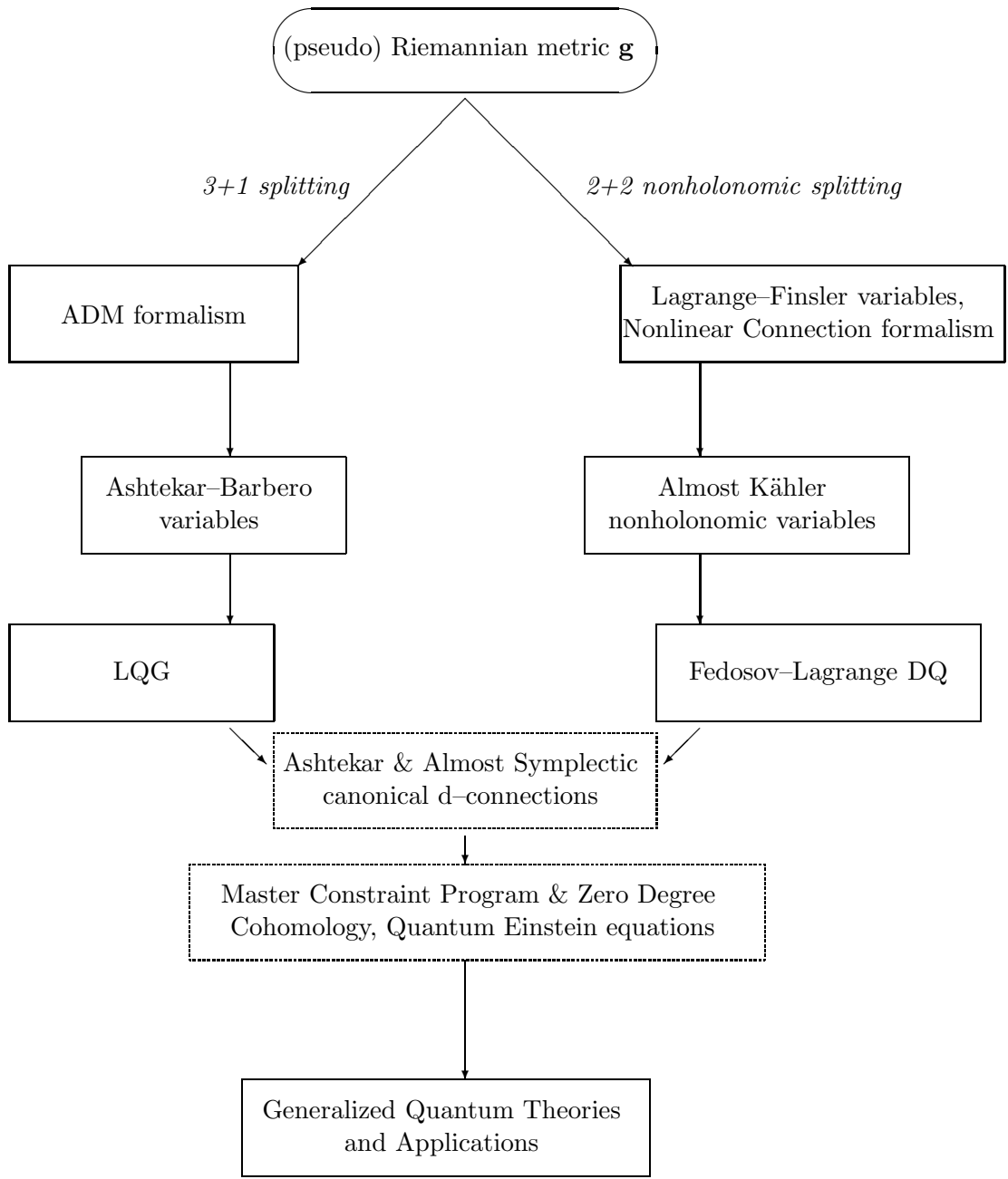

A formal scheme for LQG and DQ of GR, sketching a mathematical physics
"dictionary" between two, in general, different nonlinear quantum theories,
is given in Figure \ref{fig1}. We see that the data for one type of
quantization can be transformed into the data for the second type
quantization, and inversely, if we generalize the Ashtekar--Barbero
connection to a nonholonomic version of linear connection adapted to a
canonical nonlinear connection structure (for the equivalent almost K\"{a}%
hler--Lagrange model). This allows us to provide a Fedosov quantization of
GR in terms of effective Lagrange--Finsler variables re--expressed as
generalized Ashtekar variables and to define the LQG geometric objects in
terms of DQ objects, and inversely.  Here, it should be emphasized that LQG gravity is
dynamical (loop quantum field) theory of spacetime but DQ of GR limited to a
definition of  the corresponding cohomology class of star products is a
geometric model of quantum mechanics for gravitational fields. Both approaches
can be developed by using different 3+1 and 2+2 splitting (including double fibrations),
preserving diffeomorphism invariance and non--perturbative character but they can not
be made equivalent like the quantum mechanics is not equivalent to quantum field theory.

We note some possible important generalizations and applications of the
unified approach to LQG and DQ proposed in this paper. Having elaborated a
quantization scheme using nonholonomic versions of Ashtekar-Barbero
variables, we can consider it in various (non) commutative gauge gravity
theories and nonlinear models of physical interactions \cite{vncg} and
quantum almost K\"{a}hler geometries related to string gravity. Inversely,
in \cite{vlqgdq2}, we show how we can quantize nonlinear mechanics models
and related Lagrange--Finsler geometries following methods LQG which gives
us the fist example of quantum analogous gravity self--consistently
formulated in the language of geometric mechanics.

Finally, we note that in this work we only sketched the proofs of the most
important results of common interest in LQG and DQ of GR since we wonted to
reach a rather general audience and find certain important common points of
these two different approaches. All the technical details can be found in
comprehensive and self--contained forms in the cited monographs and reviews.

\vskip3pt

\textbf{Acknowledgement: } This work was elaborated during a long term author's visit at Fields
Institute and short term visit at Perimeter Institute. He is very obliged to J. Moffat,
for kind support and hosting, and L. Smolin and L. Freidel, for important critical remarks and
discussions.

\appendix

\setcounter{equation}{0} \renewcommand{\theequation}
{A.\arabic{equation}} \setcounter{subsection}{0}
\renewcommand{\thesubsection}
{A.\arabic{subsection}}

\section{Almost complex adapted Ashtekar Connections}

The coefficients of the Neijenhuis tensor $\ ^{\mathbf{J}}\mathbf{\Omega }$ (%
\ref{neijt}) for the canonical almost complex structure $\mathbf{J(\ ^{L}%
\mathbf{g})}$ (\ref{acstr}) are
\begin{equation}
\mathbf{\ }^{\mathbf{J}}\mathbf{\Omega }_{\alpha \beta }^{\gamma }=(\mathbf{e%
}_{\tau }\mathbf{J}_{\ \alpha }^{\gamma })\mathbf{J}_{\ \beta }^{\tau }-(%
\mathbf{e}_{\tau }\mathbf{J}_{\ \beta }^{\gamma })\mathbf{J}_{\ \alpha
}^{\tau }+(\mathbf{e}_{\alpha }\mathbf{J}_{\ \beta }^{\tau }-\mathbf{e}%
_{\beta }\mathbf{J}_{\ \alpha }^{\tau })\mathbf{J}_{\ \tau }^{\gamma },
\label{cnjt}
\end{equation}%
where $\mathbf{J}(\mathbf{e}_{\beta })=\mathbf{J}_{\ \beta }^{\alpha }%
\mathbf{e}_{\alpha }$ is decomposed with respect to N--adapted frames (\ref%
{dder}) defined by the canonical N--connection structure. Let us consider a
d--connection
\begin{eqnarray}
\ ^{A}\mathbf{\Gamma }_{\beta \gamma }^{\alpha }\mathbf{(\ \mathbf{g})} &=&\
\ ^{c}\mathbf{\Gamma }_{\beta \gamma }^{\alpha }\mathbf{(\mathbf{g})}+\ \
^{c}\mathbf{Z}_{\beta \gamma }^{\alpha }\mathbf{(\mathbf{g})}
\label{defashtd} \\
&=&\ \widehat{\Gamma }_{\beta \gamma }^{\alpha }\mathbf{(\mathbf{g})}+\
\widehat{\mathbf{Z}}_{\beta \gamma }^{\alpha }\mathbf{(\mathbf{g})}=\
_{\shortmid }\Gamma _{\beta \gamma }^{\alpha }\mathbf{(\mathbf{g})}+\
_{\shortmid }^{A}\mathbf{Z}_{\beta \gamma }^{\alpha }\mathbf{(\mathbf{g})},
\notag
\end{eqnarray}%
with deformation formulas similar to $\ _{\shortmid }\Gamma _{\ \alpha \beta
}^{\gamma }\mathbf{(\mathbf{g})}=\widehat{\mathbf{\Gamma }}_{\ \alpha \beta
}^{\gamma }\mathbf{(\mathbf{g})}+\ _{\shortmid }Z_{\ \alpha \beta }^{\gamma }%
\mathbf{(\mathbf{g})}$ (\ref{cdeft}), where $\ _{\shortmid }^{A}\mathbf{Z}%
_{\beta \gamma }^{\alpha }$ is just the distorsion of the Levi--Civita
connection $\ _{\shortmid }\Gamma _{\beta \gamma }^{\alpha }$ defining the
connection $^{A}\mathbf{\Gamma }_{\beta \gamma }^{\alpha }$ with torsion $\
^{A}\mathbf{T}_{\beta \gamma }^{\alpha }=-\mathbf{\ }^{\mathbf{J}}\mathbf{%
\Omega }_{\alpha \beta }^{\gamma }/4.$ All linear connections, $^{A}\mathbf{%
\Gamma ,}\ \ \ ^{c}\mathbf{\Gamma ,}\ \widehat{\Gamma }$ and $\ \
_{\shortmid }\Gamma ,$ and deformation tensors, $\ ^{c}\mathbf{Z,}\ \widehat{%
\mathbf{Z}}$ and $\ _{\shortmid }^{A}\mathbf{Z,}$ are uniquely defined by
the metric tensor $\mathbf{g=\ ^{L}g}$ on $\mathbf{V}.$ Having $\
_{\shortmid }^{A}\mathbf{Z}$ determined by $\mathbf{\ }^{\mathbf{J}}\mathbf{%
\Omega ,}$ we compute the distorsion formulas relating $\ ^{c}\mathbf{\Gamma
}$ and $\ _{\shortmid }\Gamma ,$ see details in \cite{vrfg,vsgg,ma}.

For any metric structure $\mathbf{g}$ on a manifold $\mathbf{V,}$ the
Levi--Civita connection is by definition the unique linear connection $%
\bigtriangledown =\{\ _{\shortmid }\Gamma _{\beta \gamma }^{\alpha }\}$
which is metric compatible, $\bigtriangledown \mathbf{g}=0,$ and torsionless,%
$~\ _{\shortmid }\mathcal{T}=0.$ This is not a d--connection because it does
not preserve under parallelism the N--connection splitting (\ref{whitney}).
We parametrize the coefficients:
\begin{equation*}
\bigtriangledown = \{\ _{\shortmid }\Gamma _{\beta \gamma }^{\alpha }=\left(
_{\shortmid }L_{jk}^{i},_{\shortmid }L_{jk}^{a},_{\shortmid }L_{bk}^{i},\
_{\shortmid }L_{bk}^{a},_{\shortmid }C_{jb}^{i},_{\shortmid
}C_{jb}^{a},_{\shortmid }C_{bc}^{i},_{\shortmid }C_{bc}^{a}\right)\},
\mbox{
where }
\end{equation*}%
\begin{eqnarray*}
\bigtriangledown _{\mathbf{e}_{k}}(\mathbf{e}_{j}) &=&\ _{\shortmid
}L_{jk}^{i}\mathbf{e}_{i}+\ _{\shortmid }L_{jk}^{a}e_{a},\ \bigtriangledown
_{\mathbf{e}_{k}}(e_{b})=\ _{\shortmid }L_{bk}^{i}\mathbf{e}_{i}+\
_{\shortmid }L_{bk}^{a}e_{a}, \\
\bigtriangledown _{e_{b}}(\mathbf{e}_{j}) &=&\ _{\shortmid }C_{jb}^{i}%
\mathbf{e}_{i}+\ _{\shortmid }C_{jb}^{a}e_{a},\ \bigtriangledown
_{e_{c}}(e_{b})=\ _{\shortmid }C_{bc}^{i}\mathbf{e}_{i}+\ _{\shortmid
}C_{bc}^{a}e_{a}.
\end{eqnarray*}%
The canonical d--connection $\ \ ^{c}\mathbf{\Gamma }_{\ \alpha \beta
}^{\gamma }=\left( \widehat{L}_{jk}^{i},\widehat{L}_{bk}^{a},\widehat{C}%
_{jc}^{i},\widehat{C}_{bc}^{a}\right) $ with coefficients
\begin{eqnarray}
\widehat{L}_{jk}^{i} &=&\frac{1}{2}\ ^{L}g^{ir}\left( e_{k}\
^{L}g_{jr}+e_{j}\ ^{L}g_{kr}-e_{r}\ ^{L}g_{jk}\right) ,  \label{candcon} \\
\widehat{L}_{bk}^{a} &=&e_{b}(\ ^{L}N_{k}^{a})+\frac{1}{2}\ ^{L}g^{ac}\left(
e_{k}\ ^{L}g_{bc}-\ ^{L}g_{dc}\ e_{b}N_{k}^{d}-\ ^{L}g_{db}\ e_{c}\
^{L}N_{k}^{d}\right) ,  \notag \\
\widehat{C}_{jc}^{i} &=&\frac{1}{2}\ ^{L}g^{ik}e_{c}\ ^{L}g_{kj},\ \widehat{C%
}_{bc}^{a}=\frac{1}{2}\ ^{L}g^{ad}\left( e_{c}\ ^{L}g_{bd}+e_{c}\
^{L}g_{cd}-e_{d}\ ^{L}g_{bc}\right)  \notag
\end{eqnarray}%
is by definition a metric compatible d--connection uniquely defined by $%
\mathbf{g}=\ ^{L}\mathbf{g}$ (\ref{lfsm}) to satisfy the properties in $\ \
^{c}T_{\ jk}^{i}=0$ and $\ \ ^{c}T_{\ bc}^{a}=0$ but $\ \ ^{c}T_{\ ja}^{i},\
\ ^{c}T_{\ ji}^{a}$ and $\ \ ^{c}T_{\ bi}^{a}$ are not zero.

A straightforward calculus shows that the coefficients of the Levi--Civita
connection can be expressed in the form%
\begin{eqnarray}
\ _{\shortmid }L_{jk}^{i} &=&\widehat{L}_{jk}^{i},\ _{\shortmid }L_{jk}^{a}=-%
\widehat{C}_{jb}^{i}\ ^{L}g_{ik}\ ^{L}g^{ab}-\frac{1}{2}\Omega _{jk}^{a},
\label{lccon} \\
\ _{\shortmid }L_{bk}^{i} &=&\frac{1}{2}\Omega _{jk}^{c}\ ^{L}g_{cb}\
^{L}g^{ji}-\frac{1}{2}(\delta _{j}^{i}\delta _{k}^{h}-\ ^{L}g_{jk}\
^{L}g^{ih})\widehat{C}_{hb}^{j},  \notag \\
\ _{\shortmid }L_{bk}^{a} &=&\widehat{L}_{bk}^{a}+\frac{1}{2}(\delta
_{c}^{a}\delta _{d}^{b}+\ ^{L}g_{cd}\ ^{L}g^{ab})\left[ \widehat{L}%
_{bk}^{c}-e_{b}(\ ^{L}N_{k}^{c})\right] ,  \notag \\
\ _{\shortmid }C_{kb}^{i} &=&\widehat{C}_{kb}^{i}+\frac{1}{2}\check{\Omega}%
_{jk}^{a}\ ^{L}g_{cb}\ ^{L}g^{ji}+\frac{1}{2}(\delta _{j}^{i}\delta
_{k}^{h}-\ ^{L}g_{jk}\ ^{L}g^{ih})\widehat{C}_{hb}^{j},  \notag \\
\ _{\shortmid }C_{jb}^{a} &=&-\frac{1}{2}(\delta _{c}^{a}\delta _{b}^{d}-\
^{L}g_{cb}\ ^{L}g^{ad})\left[ \widehat{L}_{dj}^{c}-e_{d}(\ ^{L}N_{j}^{c})%
\right] ,\ _{\shortmid }C_{bc}^{a}=\widehat{C}_{bc}^{a},  \notag \\
\ _{\shortmid }C_{ab}^{i} &=&-\frac{\ ^{L}g^{ij}}{2}\left\{ \left[ \widehat{L%
}_{aj}^{c}-e_{a}(\ ^{L}N_{j}^{c})\right] \ ^{L}g_{cb}+\left[ \widehat{L}%
_{bj}^{c}-e_{b}(\ ^{L}N_{j}^{c})\right] \ ^{L}g_{ca}\right\} ,  \notag
\end{eqnarray}%
For spacetimes of even dimension, instead of $\ ^{c}\mathbf{\Gamma }_{\
\alpha \beta }^{\gamma },$ we can consider the normal d--connection $%
\widehat{\mathbf{\Gamma }}_{\beta \gamma }^{\alpha }=(\widehat{L}_{jk}^{i},\
^{v}\widehat{C}_{bc}^{a})$ (\ref{ndc}) with coefficients (\ref{cdcc}).

Let introduce the distorsion d--tensor $\ _{\shortmid }Z_{\ \alpha \beta
}^{\gamma }$ with N--adapted coefficients%
\begin{eqnarray}
\ _{\shortmid }Z_{jk}^{i} &=&0,\ _{\shortmid }Z_{jk}^{a}=-\widehat{C}%
_{jb}^{i}\ ^{L}g_{ik}\ ^{L}g^{ab}-\frac{1}{2}\Omega _{jk}^{a},  \notag \\
\ _{\shortmid }Z_{bk}^{i} &=&\frac{1}{2}\Omega _{jk}^{c}\ ^{L}g_{cb}\
^{L}g^{ji}-\frac{1}{2}(\delta _{j}^{i}\delta _{k}^{h}-\ ^{L}g_{jk}\
^{L}g^{ih})\widehat{C}_{hb}^{j},  \notag \\
\ _{\shortmid }Z_{bk}^{a} &=&\frac{1}{2}(\delta _{c}^{a}\delta _{d}^{b}+\
^{L}g_{cd}\ ^{L}g^{ab})\left[ \widehat{L}_{bk}^{c}-e_{b}(\ ^{L}N_{k}^{c})%
\right] ,  \notag \\
\ _{\shortmid }Z_{kb}^{i} &=&\frac{1}{2}\Omega _{jk}^{a}\ ^{L}g_{ab}\
^{L}g^{ji}+\frac{1}{2}(\delta _{j}^{i}\delta _{k}^{h}-\ ^{L}g_{jk}\
^{L}g^{ih})\widehat{C}_{hb}^{j},  \notag \\
\ _{\shortmid }Z_{jb}^{a} &=&-\frac{1}{2}(\delta _{c}^{a}\delta _{b}^{d}-\
^{L}g_{cb}\ ^{L}g^{ad})\left[ \widehat{L}_{dj}^{c}-e_{d}(\ ^{L}N_{j}^{c})%
\right] ,\ _{\shortmid }Z_{bc}^{a}=0,  \label{cdeftc} \\
\ _{\shortmid }Z_{ab}^{i} &=&-\frac{\ ^{L}g^{ij}}{2}\left\{ \left[ \widehat{L%
}_{aj}^{c}-e_{a}(\ ^{L}N_{j}^{c})\right] \check{g}_{cb}+\left[ \widehat{L}%
_{bj}^{c}-e_{b}(\ ^{L}N_{j}^{c})\right] \ ^{L}g_{ca}\right\} .  \notag
\end{eqnarray}

In the simplest case, having computed (\ref{neijt}), following relations (%
\ref{defashtd}), (\ref{candcon}) and (\ref{lccon}), we express $\
_{\shortmid }^{A}\mathbf{Z}_{\beta \gamma }^{\alpha }\mathbf{(\mathbf{g})}$
from (\ref{defashtd}) as%
\begin{equation*}
\ _{\shortmid }^{A}\mathbf{Z}_{\beta \gamma }^{\alpha }=-\frac{1}{8}\mathbf{%
\mathbf{g}}^{\alpha \tau }\left( \mathbf{\mathbf{g}}_{\gamma \varepsilon }%
\mathbf{\ }^{\mathbf{J}}\mathbf{\Omega }_{\tau \beta }^{\varepsilon }\mathbf{%
\ +}\ \mathbf{\mathbf{g}}_{\beta \varepsilon }\mathbf{\ }^{\mathbf{J}}%
\mathbf{\Omega }_{\tau \gamma }^{\varepsilon }-\ \mathbf{\mathbf{g}}_{\tau
\varepsilon }\mathbf{\ }^{\mathbf{J}}\mathbf{\Omega }_{\beta \gamma
}^{\varepsilon }\right) ,
\end{equation*}%
which defines completely $\ ^{A}\mathbf{\Gamma }_{\beta \gamma }^{\alpha }%
\mathbf{(\mathbf{g}).}$ \ At the final step, we consider the formulas (\ref%
{so3cf}) but for the d--connection $\ ^{A}\mathbf{\Gamma ,}$
\begin{equation}
\ \ ^{A}\mathbf{\Gamma }_{\ I}^{I^{\prime }}\doteqdot \frac{1}{2}%
q_{I}^{\alpha }q_{\alpha ^{\prime }}^{I^{\prime }}\epsilon _{\quad \gamma
^{\prime }\tau ^{\prime }}^{\alpha ^{\prime }\beta ^{\prime }}\mathbf{n}%
_{\beta ^{\prime }}\ \ ^{A}\mathbf{\Gamma }_{\quad \alpha }^{\gamma ^{\prime
}\tau ^{\prime }}\mbox{ and }\ ^{A}\mathbf{K}_{\ I}^{I^{\prime }}\doteqdot
q_{\alpha ^{\prime }}^{I^{\prime }}q_{I}^{\alpha }\mathbf{n}_{\beta ^{\prime
}}\ \ ^{A}\mathbf{\Gamma }_{\quad \alpha }^{\alpha ^{\prime }\beta ^{\prime
}},  \label{ashtaux}
\end{equation}%
and define the Ashtekar d--connection $\mathbf{A}_{\ I}^{I^{\prime }}=\
_{\shortmid }A_{\ I}^{I^{\prime }}+\ _{\shortmid }^{A}\mathbf{Z}_{\
I}^{I^{\prime }}$ (\ref{ndashc}), where
\begin{equation}
\ \ \ _{\shortmid }^{A}\mathbf{Z}_{\ I}^{I^{\prime }}\doteqdot q_{I}^{\alpha
}q_{\alpha ^{\prime }}^{I^{\prime }}\mathbf{n}_{\beta ^{\prime }}\left(
\frac{1}{2}\epsilon _{\quad \gamma ^{\prime }\tau ^{\prime }}^{\alpha
^{\prime }\beta ^{\prime }}\ \ \ _{\shortmid }^{A}\mathbf{Z}_{\quad \alpha
}^{\gamma ^{\prime }\tau ^{\prime }}+\beta \ \ _{\shortmid }^{A}\mathbf{Z}%
_{\quad \alpha }^{\alpha ^{\prime }\beta ^{\prime }}\right)  \label{ashtdt}
\end{equation}%
and $\ _{\shortmid }A_{\ I}^{I^{\prime }}$ are given by formulas (\ref%
{ashtcon}).

\end{document}